\documentclass[12pt,notitlepage,tightenlines,eqsecnum,floats,nofootinbib,amsmath,amssymb,aps,prd]{revtex4-1}
\usepackage{amsmath,amssymb,amsfonts,latexsym,cancel}

\usepackage{mathtools}
\usepackage{graphicx}
\usepackage{color}
\usepackage{amsbsy}
\usepackage{hyperref}
\usepackage{tikz}

\usepackage{comment}

\newcommand{\be}{\begin{equation}}
\newcommand{\beq}{\begin{equation}}
\newcommand{\ee}{\end{equation}}

\newcommand{\braket}[1]{\langle #1 \rangle}

\newcommand{\ket}[1]{| #1 \rangle}

\newcommand{\abs}[1]{\left\lvert#1\right\rvert}

\def\bea {\begin{eqnarray}}
\def\eea {\end{eqnarray}}

\newcommand{\pd}{\partial}

\def\ea{{E}^{a}}
\def\eb{{E}^{b}}

\def\f{\frac}
\def\nn{\nonumber}
\def\lp{\ell_{\rm Pl}}

\def\b{\bar}

\def\h{\hat}
\def\t{\tilde}

\def\dd{{\rm d}}

\def\del{\partial}

\def\de{\delta}

\def\ga{\gamma}

\def\De{\Delta}

\def\mH{\mathcal{H}}

\def\mN{\mathcal{N}}

\def\mE{\mathcal{E}}

\begin{document}

\title{On the fate of quantum black holes}

\author{Viqar Husain} \email{vhusain@unb.ca}
\affiliation{Department of Mathematics and Statistics, University of New Brunswick,
Fredericton, NB, Canada E3B 5A3}

\author{Jarod George Kelly} \email{jarod.kelly@unb.ca}
\affiliation{Department of Mathematics and Statistics, University of New Brunswick,
Fredericton, NB, Canada E3B 5A3}

\author{Robert Santacruz} \email{robert.santacruz@unb.ca}
\affiliation{Department of Mathematics and Statistics, University of New Brunswick,
Fredericton, NB, Canada E3B 5A3}

\author{Edward Wilson-Ewing} \email{edward.wilson-ewing@unb.ca}
\affiliation{Department of Mathematics and Statistics, University of New Brunswick,
Fredericton, NB, Canada E3B 5A3}

\begin{abstract}

We study the quantum dynamics of the Lema\^itre-Tolman-Bondi space-times using a polymer quantization prescription based on loop quantum cosmology that incorporates fundamental discreteness. By solving an effective equation derived from this quantization, we find analytical solutions for the Oppenheimer-Snyder and thin-shell collapse models, and numerical solutions for a variety of asymptotically flat collapsing dust profiles. Our study (i) tracks the formation, evolution and disappearance of dynamical horizons, (ii) shows that matter undergoes a non-singular bounce that results in an outgoing shock wave, (iii) determines black hole lifetime to be proportional to the square its mass, and (iv) provides a conformal diagram that substantially modifies the standard ``information loss" picture by resolving the singularity and replacing the event horizon by transient apparent horizons. 

\end{abstract}

\maketitle

\section{Introduction}

In general relativity, collapsing matter may form a black hole depending on the initial conditions \cite{Christodoulou:1987vu,Choptuik:1992jv}. The simplest black hole in classical general relativity is spherically symmetric, singular, and stable. In the semiclassical theory where quantum fields propagate on classical space-times, black holes are unstable due to Hawking radiation \cite{Hawking:1975vcx}. The endpoint of the process of Hawking evaporation remains unknown---it is expected that a theory of quantum gravity will answer this question by resolving the singularity and providing dynamics past the classically singular region. 

One approach to quantum gravity is loop quantum gravity (LQG), a non-perturbative attempt to quantize general relativity using a background-independent Hilbert space where the inner product does not utilize the space-time metric \cite{Ashtekar:2004eh, Rovelli:2004tv, Thiemann:2007pyv}. Solving the full quantum dynamics in LQG currently remains out of reach, but there is a substantial body of work on symmetry-reduced models, including applications to homogeneous cosmological space-times known as loop quantum cosmology (LQC) \cite{Ashtekar:2011ni}, and to inhomogeneous spherically symmetric models including black hole space-times \cite{Perez:2017cmj}.

There are two main inputs from LQG in studying these simpler systems, the holonomy and inverse triad operators. These arise due to the quantization procedure used: (i) the connection and its curvature do not exist as operators and must be defined using the holonomy operators, and (ii) a well-defined inverse triad operator is required for constructing the constraint operators. These two features lead to what are respectively called ``holonomy" and ``inverse triad" corrections. 

In symmetry-reduced models these features arise through polymer quantization. Denoting the phase space configuration variables $q_j$ (corresponding to the connection) and $p_j$ (corresponding to spatial metric degrees of freedom), the polymer quantization prescription is based on elementary operators corresponding to the (non-differentiable) complex exponential $\hat{U}_\mu(q_j)=\widehat{\exp(i\mu q_j)}$ and the operator $\hat p_j$; (the inverse momentum operators $\widehat{p_j^{-1}}$ corresponding to the inverse triad operator of LQG is defined using these elementray operators). The parameter $\mu$ in the operator $\hat{U}_\mu(q_j)$ corresponds to the coordinate length of a holonomy segment. This parameter is chosen by requiring that the holonomies appearing in curvature operators must have a physical length of the order of the Planck length $\lp$. This requirement is implemented by relating the coordinate length $\mu$ to a physical length using the spatial metric. Since the spatial metric can be reconstructed from $p_j$ it follows that $\mu=\mu(p_j)$; this prescription is known in the literature (for historical reasons) as the $\bar\mu$ scheme or ``improved dynamics".

In LQC, these methods have successfully led to the resolution of the big-bang singularity in classical general relativity, replacing it with a non-singular bounce in homogeneous cosmological space-times \cite{Ashtekar:2006wn, Ashtekar:2011ni}. 

The application of LQG techniques to black hole space-times falls into three main categories. Works in the first category use the classical isometry between the black hole interior and the Kantowski-Sachs homogeneous space-time. This suggests a description of the black hole interior where LQC methods can be applied directly \cite{Modesto:2004xx, Ashtekar:2005qt, Bohmer:2007wi, Campiglia:2007pb, Chiou:2008nm, Brannlund:2008iw, Joe:2014tca, Corichi:2015xia, Cortez:2017alh, Olmedo:2017lvt, BenAchour:2018khr, Ashtekar:2018cay, Bodendorfer:2019cyv, Assanioussi:2019twp, Bodendorfer:2019nvy, Zhang:2020qxw, Sartini:2020ycs, Geiller:2020xze, Blanchette:2020kkk}. Although simple, this approach has several drawbacks: the isometry between the Schwarzschild interior and the Kantowski-Sachs space-time may not hold in full quantum gravity (it is known to fail for some modified gravity theories \cite{deCesare:2020swb}); the isometry requires the presence of an outer horizon and assumes there is no inner horizon; and the standard improved dynamics scheme as applied in \cite{Bohmer:2007wi} fails near the horizon (likely due to the spatial coordinates becoming null).

In the second category are works that use coordinates in spherical symmetry where the radial coordinate remains spacelike throughout the entire space-time, making it is possible to treat the interior and exterior of the black hole on an equal footing for quantization \cite{Bojowald:2005cb, Campiglia:2007pr, Gambini:2008dy, Reyes:2009, Bojowald:2011js, Gambini:2013hna, BenAchour:2016brs, Bojowald:2018xxu, Alonso-Bardaji:2021tvy}. Building on earlier works \cite{Bohmer:2007wi, Chiou:2012pg}, the improved dynamics scheme was recently applied to vacuum spherically symmetric space-times \cite{Gambini:2020nsf, Kelly:2020uwj, Han:2020uhb}. These works provide a framework that can be used to study the entire black hole space-time (not just the interior). It has the advantage that the correct classical limit is recovered where the space-time curvature is small compared to the Planck scale. 

There is also work in a third category which aims for a full LQG treatment of states potentially corresponding to black holes in canonical LQG \cite{Thiemann:1992jj, Alesci:2019pbs, Alesci:2020zfi}, in the covariant spin foam approach \cite{Christodoulou:2016vny, Christodoulou:2018ryl, Bianchi:2018mml, Martin-Dussaud:2019wqc, DAmbrosio:2020mut, Soltani:2021zmv}, and in the group field theory reformulation of LQG \cite{Oriti:2018qty}. Although it is challenging to include inhomogeneous matter fields and make contact with semiclassical physics, these works provide some guidance for recovering low curvature classical black hole space-times from full quantum gravity.

To determine the fate of a black hole in quantum gravity, it is important to include matter so that the entire process from gravitational collapse to black hole formation to its entire subsequent evolution is captured. Initial steps in this direction used Oppenheimer-Snyder type models where the interior is an FRLW cosmology \cite{Modesto:2006qh, Hossenfelder:2009fc, Tavakoli:2013rna, Bambi:2013caa, Liu:2014kra, BenAchour:2020bdt}, and thin shells \cite{Campiglia:2016fzp, Ziprick:2016ogy,Giesel:2021dug}. Beyond this there is work on inhomogeneous models including a scalar field with holonomy corrections \cite{Benitez:2020szx} or with inverse triad corrections \cite{Husain:2004yz, Ziprick:2009nd, Kreienbuehl:2010vc}, a dust field with inverse triad corrections \cite{Bojowald:2008ja, Bojowald:2009ih}, and Wheeler-DeWitt canonical quantization \cite{Vaz:2011zz}. 

A generic feature of these works is that, as in cosmology, the singularity is resolved and matter bounces when space-time curvature reaches the Planck scale. This is seen in LQG \cite{Modesto:2006qh, Hossenfelder:2009fc, Tavakoli:2013rna, Bambi:2013caa, Liu:2014kra, BenAchour:2020bdt} and in other approaches \cite{Barcelo:2014npa, Barcelo:2014cla, Kiefer:2019csi, Schmitz:2019jct, Piechocki:2020bfo, Schmitz:2020vdr, Munch:2020czs}; a model-independent view is studied in \cite{BenAchour:2020gon}, and a recent review is \cite{Malafarina:2017csn}.

It may also be possible to understand the bounce generated by quantum gravity effects as a transition from a black hole to a white hole \cite{Rovelli:2014cta, Haggard:2014rza}, with potential observational consequences \cite{Barrau:2014hda, Barrau:2014yka, Barrau:2015uca}. However, since white holes are known to be unstable \cite{Eardley:1974zz, Barcelo:2015uff} this scenario requires further study. An additional feature of non-singular black holes in classical gravity is mass inflation at inner horizons \cite{Poisson:1989zz, Ori:1991zz, Husain:1994xa}. We will see that the model we consider here is free of these two problems.

Building on earlier work on the vacuum Schwarzschild space-time \cite{Kelly:2020uwj}, the improved dynamics scheme was applied to the entire space-time of a spherically symmetric inhomogeneous dust model, and used to study the quantization of the Oppenheimer-Snyder model \cite{Kelly:2020lec}. It was found that quantum gravity effects halt the collapse, resolve the singularity and cause a bounce; the dust then expands outwards until the outer horizon disappears. This led to an estimate of $M^2 / m_{\rm Pl}$ for the lifetime of this quantum black hole.

In this paper, and the companion letter \cite{Husain:2021ojz}, we revisit this inhomogeneous dust model to define the quantum theory in full. We then derive the effective equations with quantum corrections for the theory; these are two coupled partial differential equations. We numerically solve the effective equations to a high degree of accuracy for a variety of initial conditions. Our solutions extend all previous works by describing evolution from initial collapse to the formation of apparent horizons, to their subsequent disappearance following a bounce and re-expansion of matter. This provides a complete lifetime scenario of black holes with matter. As further developments to the companion letter \cite{Husain:2021ojz}, this paper contains (i) significant calculations concerning the quantum theory and the derivation of the effective dynamics, (ii) analytic results for Oppenheimer-Snyder and thin shell solutions, (iii) an in-depth description of the main features in the conformal diagram, and (iv) a discussion on some implications for the information loss problem and possible astrophysical consequences.

An important feature and common point of concern for dust collapse models in classical gravity is the formation of caustics, also referred to as ``shell-crossing singularities." This feature is often viewed as undesirable; it can be avoided in the classical theory by restricting the initial data so the central singularity is reached before caustic formation \cite{Szekeres:1995gy, Hellaby:1985zz, Booth:2005ng, Giesel:2009jp}. The same restriction may also be imposed in quantized models \cite{Giesel:2021dug}; however, such restrictions are undesirable as they limit the exploration of possible solutions.

In a quantized theory that has the potential to resolve singularities, caustics are unavoidable: ingoing matter trajectories will inevitably cross outgoing ones during matter evolution past a bounce. For this reason it is essential to avoid excising potential caustics ``by hand" and instead permit so-called ``weak solutions" \cite{Nolan:2003wp, Lasky:2006hq}. Such solutions are an essential part of non-linear wave dynamics. A common feature of weak solutions is a shock wave, corresponding to a discontinuity in the field. In the gravitational case, the shock wave corresponds to a discontinuity in the metric, and therefore in the geometry of space-time.

The effective equation we derive and solve numerically is a non-linear $1+1$ dimensional continuity equation. We solve this equation numerically using the well-known Godunov method \cite{Leveque}; this is one of a class of so-called ``finite volume" methods for solving non-linear continuity equations. This method generates weak numerical solutions past the points of characteristic crossing, resulting in shock wave formation. Using this scheme we simulate the collapse of initial matter configurations where no horizon is initially present. We find generically that as the dust profile falls inward, a pair of apparent horizons form when the Schwarzschild radius is reached. The collapse continues until the space-time curvature reaches the Planck scale, and matter bounces outward as a shock wave; the shock is a discontinuity in the metric. The shock wave moves outward together with the inner apparent horizon until they reach the outer horizon located at the Schwarzschild radius; at this point both apparent horizons merge and disappear. To an outside observer this would appear as a spherical shock wave emanating from a horizon. 

Our numerical procedure allows a calculation of the black hole lifetime, defined as the time between the formation and disappearance of the outermost apparent horizon as measured by a distant observer. After evolutions spanning two orders of magnitude in data mass $M$, we find that this lifetime is $T\sim M^2 / m_{\rm Pl}$. This extends and reinforces the estimate obtained for Oppenheimer-Snyder collapse that yielded the same result \cite{Kelly:2020lec}. Furthermore, from the evolution of a double peaked Gaussian data, we demonstrate that a collapsing pulse followed by another does not result in recollapse, and does not cause mass inflation at the inner horizons; in all cases the final state is a single outgoing shock wave.

The structure of the paper is as follows: in Sec.~\ref{s.prelim} we review the classical theory of Lema\^itre-Tolman-Bondi space-times; in Sec.~\ref{s.quantum} we define the quantum theory; in Sec.~\ref{s.effective} we derive the effective dynamics from the quantum theory, consider some simple cases and determine the conformal diagram for dust collapse in this model; in Sec.~\ref{s.numerics} we summarize the numerical method we use and present our results; in Sec.~\ref{s.impl} we discuss potential implications of our work including some consequences for the information loss problem; and we conclude in Sec.~\ref{s.discussion} with a summary and outlook.

\section{Classical theory}
\label{s.prelim}

In this section we review the class of Lema\^itre-Tolman-Bondi (LTB) metrics and its Hamiltonian theory in the connection-triad variables as a prelude to studying the quantum theory.

\subsection{Hamiltonian Lema\^itre-Tolman-Bondi dynamics}
\label{s.ltb}

LTB space-times are spherically symmetric solutions of Einstein gravity with a pressureless dust field that provide simple models for stellar collapse and non-linear cosmological perturbations. The metric is often written in the diagonal form \cite{Bolejko:2011jc}
\be
\label{eq:ltbmetric1}
\dd s^2 = -\dd t^2 + \f{(\pd_r R(r,t))^2}{1+\mE(r)} \dd r^2 + R(r,t)^2 \dd\Omega^2,
\ee
where $\dd\Omega^2 = \dd\theta^2 + \sin^2 \theta \, \dd\phi^2$. The dust stress energy tensor is $T_{\mu\nu} = \rho(x,t) \, u_{\mu} u_{\nu}$, where $u_\mu = \partial_\mu T$ is the 4-velocity of the dust field and $T$ is the dust field variable.

Although the coordinate system \eqref{eq:ltbmetric1} is commonly used, a different set of coordinates based on a generalization of the Painlev\'e-Gullstrand coordinates of the Schwarzschild space-time is better suited for a Hamiltonian description of the dynamics \cite{Lasky:2006hq}. Introducing the coordinate $x=R(r,t)$, the metric becomes
\beq
\label{eq:admltb}
\dd s^2 = - \dd t^2 + \f{1}{1 + \mE(x,t)} \left( \dd x + N^x \dd t \right)^2 + x^2 \dd \Omega^2,
\ee
with $N^x = - \pd_t R$. The Schwarzschild solution in Painlev\'e-Gullstrand coordinates is recovered in the vacuum $T_{\mu\nu} = 0$ case \cite{Lasky:2006hq}.

For the quantum theory it is convenient to use the connection-triad Hamiltonian formulation minimally coupled to dust. In spherical symmetry, the densitized triads have two independent components corresponding to the radial and angular directions,
\beq
E^x_1 = E^a \sin\theta, \qquad E^\theta_2 = E^b \sin\theta, \qquad E^\phi_3 = E^b.
\ee
The canonically conjugate Ashtekar-Barbero connection $A^i_a = \Gamma^i_a + \gamma K^i_a$ is defined in terms of the spin-connection $\Gamma_a^i$ and the extrinsic curvature $K_a^i$, and $\gamma$ is the Barbero-Immirzi parameter \cite{Immirzi:1996dr, BarberoG:1994eia} (note that the classical theory is independent of the value of $\gamma$, but not the quantum theory). In spherical symmetry, the Ashtekar-Barbero connection can be parametrized as
\beq
A^i_a \tau_i dx^a = a \, \tau_1 \, dx + \left( b \, \tau_2 - \f{\partial_x \ea}{2 \eb} \, \tau_3 \right) d\theta + \left( -\cot\theta \tau_1 + \f{\partial_x \ea}{2 \eb} \tau_2 + b \, \tau_3 \right) \sin\theta \, d\phi.
\ee

As a function of densitized triads, the spherically symmetric metrics take the form 
\be
\label{eq:gen-metric}
\dd s^2 = -N^2 \dd \tau^2 + \f{(\eb)^2}{\ea} (\dd x+N^x \dd t)^2 + \ea \dd\Omega^2,
\ee
where $N(x,t)$ and $N^x(x,t)$ are the lapse function and radial shift vector.

In the general case $E^a, E^b, a, b, N$ and $N^x$, as well as $T$ and its momentum $p_T$, are all functions of $x$ and $t$. To make contact with the metric \eqref{eq:admltb} we impose the radial gauge fixing $E^a = x^2$, which sets $N^x = -b/\gamma$ to ensure the gauge is preserved by the dynamics \cite{Kelly:2020uwj}. Additionally it is convenient to fix the dust time gauge $T=t$ which requires $N=1$ as the gauge-fixing condition \cite{Husain:2011tk}. With these gauge choices that fix $a, E^a, T, p_T$, there remains only a two-dimensional phase space parametrized by $(b,\eb)$, and the metric \eqref{eq:gen-metric} exactly matches the form given in \eqref{eq:admltb},
\be
\label{eq:metric-gf}
\dd s^2 = - \dd t^2 + \f{(\eb)^2}{x^2} \left( \dd x - \f{b}{\gamma} \, \dd t \right)^2 + x^2 \dd\Omega^2,
\ee
with $(E^b)^2/x^2 = 1/(1+\mE)$.

Since all gauge freedom is fixed at this stage, the constraints have been solved and there only remains the physical canonical action \cite{Kelly:2020uwj, Kelly:2020lec}
\beq
\label{eq:redH}
S_{GF} = \int \dd t \int \dd x \, \left( \f{\dot{b}\eb}{G\gamma} - \mathcal{H}_{\rm phys} \right),
\ee
where 
\beq
\mathcal{H}_{\rm phys} = - \, \frac{1}{2G\gamma} \biggr[ \frac{\abs{\eb}}{\gamma x} \pd_x( x b^2 ) + \frac{\gamma \abs{\eb}}{x} 
+ \frac{2\gamma x^2}{(\eb)^2} \pd_x \abs{\eb} - \frac{3 \gamma x}{\abs{\eb}} \biggr]
\ee
is the physical Hamiltonian that generates evolution with respect to physical time $T$ (which has been gauge-fixed to $T=t$). The third and fourth terms can be combined by integrating by parts the third term, keeping the total derivative the result is
\beq
\mH_{\rm phys} = - \, \frac{1}{2G\gamma} \left[ \frac{\abs{\eb}}{\gamma x} \pd_x( x b^2 ) + \frac{\gamma \abs{\eb}}{x} + \frac{\gamma x}{\abs{\eb}} - 2 \gamma \, \pd_x \left( \f{x^2}{|\eb|} \right) \right].
\ee
The total derivative does not affect the equations of motion, but it is important when calculating the energy density of the dust through $\mH_{\rm phys}$, see below.

The gauge-fixed action \eqref{eq:redH} also determines the symplectic structure for the remaining phase space variables,
\beq
\{b(x_1), E^b(x_2)\} = G \ga ~ \delta(x_1 - x_2).
\ee

The dust energy density is related to the physical Hamiltonian by \cite{Kelly:2020lec}
\be \label{density}
\rho = - ~ \f{\mathcal{H}_{\rm phys}}{4 \pi x \, \abs{\eb}},
\ee
and the total mass contained within a radius $x$ is 
\beq
m(x,t) = 4\pi \int_0^x \! \dd \tilde x \, \tilde x \, \abs{E^b} \rho(\tilde{x},t) = - \int_0^x \! \dd \t x \, \mH_{\rm phys}(\tilde{x},t).
\ee
The equations of motion follow from $\dot f = \{f, \int \! dx ~ \mathcal{H}_{\rm phys} \}$, assuming $E^b > 0$ they are
\begin{align}
\dot \eb & = \f{b \eb}{\gamma x} - \f{b}{\gamma} \, \pd_x \eb, \\
\dot b &= \f{\gamma x}{2 (\eb)^2} - \f{1}{2 \gamma x} \Big( 2x b \pd_x b + b^2 + \gamma^2 \Big).
\end{align}

\subsection{Marginally-trapped solutions}
\label{sec:mts}

A particularly simple family of LTB metrics are those with $\mE=0$ in the metric \eqref{eq:admltb}, referred to in the literature as marginally trapped solutions. Physically these correspond to solutions where the energy of all dust particles is zero (with their positive kinetic energy cancelled by their negative gravitational potential energy, and both going to zero at infinity). There are many other interesting families of LTB metrics with $\mE \neq 0$, but as a first step in studying LTB space-times in effective LQG we focus here on the marginally trapped class of solutions. In triad variables these correspond to $E^b=x$, which is readily seen as a solution to the $E^b$ equation of motion, and the remaining canonical equation for $b(x,t)$ simplifies to
\be
\label{eq:redb}
\dot{b}+\f{1}{2\gamma x}\pd_x(xb^2)=0.
\ee
The dust energy density for this class of solutions is 
\be \label{rho-marg}
\rho(x,t) = \f{1}{8\pi G \gamma^2 x^2}\pd_x\left(xb^2\right),
\ee
and the mass function is 
\be
m(x,t) = 4\pi\int_0^x \dd \t x ~ \tilde{x}^2\rho(\tilde{x},t).
\ee
As an aside, note that the equation of motion for $b$ may also be rewritten in terms of the mass function \cite{Lasky:2006hq}
\beq
\dot m + \sqrt \f{2Gm}{x} ~ \pd_x m = 0.
\ee

The dynamics for this system can be solved by the method of characteristics by writing $b = b( x(\lambda), t(\lambda))$ to transform \eqref{eq:redb} into the equations 
\beq
\frac{\dd t}{\dd\lambda} = 1, \qquad
\frac{\dd x}{\dd\lambda} = \frac{b(\lambda)}{\gamma}, \qquad
\frac{\dd b}{\dd\lambda} = -\frac{b^2(\lambda)}{2\gamma x(\lambda)}.
\ee
The first two equations determine the characteristic curves in the $(x,t)$ plane, while the third gives the value of $b$ along these curves. These may be solved with the initial conditions 
\beq
t(\lambda=0) = 0, \qquad
x(\lambda=0) = x_0, \qquad
b(\lambda=0) = b_0(x_0),
\ee
where $b_0(x) = b(x,0)$ is the initial data for $b$; it may be determined from an initial profile for the dust energy density $\rho_0(x)$ by inverting \eqref{rho-marg}.

It is straightforward to solve these coupled ODEs: the first gives $\lambda = t$, while the equation for $\dd(x^{-1}b)/\dd\lambda$ is separable, leaving a final ODE which is also separable once $b/x$ is known. The solution for $x(\lambda)$ and $b(\lambda)$ is
\beq
x(\lambda) = x_0 \left(1+ \f{3 b_0(x_0)}{2\gamma x_0} \lambda \right)^{2/3}, \qquad
b(\lambda) = \f{b_0(x_0)}{\left( 1 + \f{3 b_0(x_0)}{2\gamma x_0} \lambda \right)^{1/3}} \, .
\ee
By following different characteristic curves (each curve labeled by its initial radial position $x_0$), it is possible to determine $b(x,t)$ by inverting $x(\lambda, x_0)$---for this system inverting $t(\lambda,x_0)$ is trivial.

Depending on the initial conditions, characteristic curves starting at different radial points $x_0$ can intersect; if this happens the solution for $b$ becomes multivalued at these points since the value of $b$ on different curves will typically differ. The space-time points where characteristic curves intersect are known as shell-crossings. Shell-crossings are characterized by the condition $\pd_{x_0} x(\lambda,x_0)=0$, these are points where the determinant of the Jacobian associated to the transformation $(\lambda, x_0) \rightarrow (t,x)$ becomes zero, and $x(\lambda, x_0)$ ceases to be invertible.

The time it takes before a shell-crossing develops depends on the initial profile $b_0(x)$, and is given by
\beq \label{shocktime}
t_S = \frac{-2 \gamma x_0}{b_0 + 2 x_0 \pd_{x_0} b_0}.
\ee
It is possible to choose initial conditions to ensure shell-crossings never occur \cite{Hellaby:1985zz}, but another approach is to extend the equations of motion to allow for weak solutions \cite{Nolan:2003wp}. This is commonly required for non-linear wave equations like \eqref{eq:redb} and leads to the observable phenomena of shock waves. For this system, the shock wave will form in the gravitational field, i.e., the metric.

\subsection{Weak solutions and shock waves}
\label{s.weak}

Weak solutions arise in many contexts, and consist of solutions to an integrated form of the equations of motion that cannot satisfy the original $n^{\rm th}$-order differential equations since they are not $n$ times differentiable. Some examples in general relativity are the Israel junction conditions for distributional matter sources such as thin shells \cite{Israel:1966}, and the Dray-'t~Hooft shock wave \cite{Dray:1984ha}.

In general, for systems described by flow equations with velocities that depend on the field, it is common for characteristic curves to intersect, leading to a breakdown of the evolution generated by the differential equations beyond the intersection point. The solution to this problem is to seek ``weak solutions" that solve an integrated form of the differential equation.

It can often happen that a weak solution is discontinuous, and then the discontinuity is known as a shock wave. Typically, shock waves propagate dynamically; a simple example is a sonic boom. Many physical phenomena exhibit shock waves, which commonly arise as weak solutions to non-linear wave equations. 

To review and illustrate the basic idea, consider the non-linear conservation equation 
\be \label{gen-wave}
\dot u + \pd_x f(u, x) = 0,
\ee
where $f(u,x)$ is a non-linear function of $u$, and potentially depends on $x$ as well.

If $u(x,t)$ satisfies the integral equation
\be
\int_0^\infty \dd t \int_{-\infty}^{\infty} \dd x \Big[(\pd_t\varphi) u + (\pd_x\varphi) f(u,x) \Big] = -\int_{-\infty}^{\infty} \dd x ~ \varphi(x,0)u(x,0)
\ee
for any test function $\varphi(x,t)$ of compact support that is continuously differentiable, then $u$ is called a weak solution of this conservation equation. If discontinuities in $u$ only arise along the $x$-axis (as would be expected for a physical system with a well-posed initial value problem), it is sufficient to consider the integral equation
\beq \label{int-eq}
\f{\dd}{\dd t} \int_{x_1}^{x_2} \dd x ~ u + \int_{x_1}^{x_2} \dd x ~ \partial_x f(u,x)
= \f{\dd}{\dd t} \int_{x_1}^{x_2} \dd x ~ u + f(u(x_2),x_2) - f(u(x_1),x_1) = 0,
\ee
for any interval $[x_1, x_2]$ of interest.

A discontinuity in $u(x)$ propagates as a shock wave with a speed determined by the integral equation. To see this, consider \eqref{int-eq} when there is a discontinuity at $x=L(t)$ in the interval $[x_1,x_2]$. Then the integral splits into two parts to give 
\beq
\f{\dd}{\dd t} \int_{x_1}^L \dd x ~ u + \f{\dd}{\dd t} \int_L^{x_2} \dd x ~ u + f(u(x_2),x_2) - f(u(x_1),x_1) = 0;
\ee
expanding the first term gives
\beq
\f{\dd L}{\dd t} \cdot \lim_{x \to L^-} u + \int_{x_1}^L \dd x ~ \pd_t u + \f{\dd L}{\dd t} \cdot \lim_{x \to L^+} u + \int_L^{x_2} \dd x ~ \pd_t u + f(u(x_2),x_2) - f(u(x_1),x_1) = 0,
\ee
where $x \to L^+$ and $x \to L^-$ denote the limits from above and below the discontinuity at $x=L$ respectively. Taking the limits $x_1 \to L^-$ and $x_2 \to L^+$, the contributions from the integrals go to 0, and solving for $\dd L / \dd t$ gives the Rankine-Hugoniot condition
\beq \label{rh-cond}
\f{\dd L}{\dd t} = \f{[f(u,x)]}{[u]},
\ee
where $[g] = \lim_{x \to L^+} g(x) - \lim_{x \to L^-} g(x)$ denotes the amplitude of the discontinuity in the function $g(x)$ at $x=L$.

Returning now to LTB space-times, we are interested in weak solutions that permit discontinuities in the metric due to shell-crossings \cite{Nolan:2003wp, Lasky:2006hq}. The equation of motion \eqref{eq:redb} can be rewritten in the form of a conservation equation with $u = xb$ and $f(u,x) = u^2 / 2 \gamma x$. It gives the integral equation
\be \label{eq:weakclas}
\f{\dd}{\dd t} \int_{x_1}^{x_2} \dd x ~ x b(x,t) + \f{x_2 b(x_2,t)^2}{2 \gamma} - \f{x_1 b(x_1,t)^2}{2 \gamma} = 0, 
\ee
and the Rankine-Hugoniot condition gives the propagation speed of a dust shock wave: 
\beq
\label{eq:cswspeed}
\f{\dd L}{\dd t} = \f{L [b^2]}{2 \gamma L[b]} = \f{b(L^+) + b(L^-)}{2\gamma}.
\ee

In general, numerics are necessary to solve the dynamics of a non-linear wave equation of the general form \eqref{gen-wave}; we discuss the Godunov numerical scheme in Sec.~\ref{s.numerics}.

Finally, we note that different prescriptions to find weak solutions are not guaranteed to be equivalent, since there are (for functions with discontinuities) many inequivalent integral forms for any given differential equation that can be obtained by non-linear field redefinitions. In particular, the weak solutions derived from \eqref{eq:weakclas} are inequivalent to the Israel junction conditions---although in both cases the induced three-dimensional metric across the discontinuous surface will be continuous, the Israel junction conditions impose an additional relation on the trace-free extrinsic curvature, while on the other hand it is a relation on $b$ that is implicitly imposed by \eqref{eq:weakclas}. We leave a more detailed exploration of weak solutions in classical general relativity for future work; in the remainder of this paper we consider weak solutions to \eqref{eq:weakclas} (and its equivalent for the LQC effective equations).

\subsection{Discretization}
\label{s.disc}

To pass to the quantum theory, it is convenient to start from a discretization of the classical theory. There are of course many discretizations that are possible and equivalent in the limit that the discretization parameter becomes infinitesimally small. Here, for the sake of simplicity we will choose a rather direct discretization.

First, we introduce a lattice in the radial direction denoted by $x_n$, with $x_0 = 0$ and the width of each interval given by $w_n = x_{n+1} - x_n > 0$. It is possible to set $w_n$ to be the same for all $n$, but that is not necessary here; we will however assume that $w_n$ is fixed for each $n$ and does not evolve dynamically.

Given this discretization, we further assume that all fields are constant in each interval $[x_n, x_{n+1})$, and approximate derivatives by
\beq
\pd_x f(x) \to \f{f(x_{n+1}) - f(x_n)}{w_n}.
\ee
With these choices, the action for the discretized theory becomes
\beq
\label{eq:redH-d}
S_{GF}^{\rm disc} = \int \dd t \sum_n \left(w_n\f{\dot{b}_n\eb_n}{G\gamma} - \mathcal{H}_n^{\rm disc} \right),
\ee
where $b_n = b(x_n), \eb_n = \eb(x_n)$ and (for $n \neq 0$)
\beq
\label{eq:discHam}
\mathcal{H}_n^{\rm disc} = - \, \frac{w_n}{2G\gamma} \biggr[ \frac{\abs{\eb_n}}{\gamma x_n} \cdot \f{x_{n+1} b_{n+1}^2 - x_n b_n^2}{w_n} 
+ \frac{\gamma \abs{\eb_n}}{x_n} 
+ \frac{\gamma x_n}{\abs{\eb_n}} \biggr],
\ee
dropping the boundary term.

Note that the same general expression cannot be used for $\mH_0^{\rm disc}$ due to the presence of $x_0=0$ in some denominators. Various regularizations to address this difficulty are possible; looking ahead to the quantum theory where there are inverse triad operators, we simply set $\mH_0^{\rm disc} = 0$.

The Poisson brackets for the discretized variables are now
\beq
\{ b_m, E^b_n \} = \f{G \ga}{w_n} ~\delta_{mn},
\ee
and the equations of motion (for $n \ge 2$ and assuming $\eb>0$) are
\be
\dot E^b_n = -\f{x_n b_n}{\gamma} \cdot \left( \f{\eb_n}{w_n x_n} - \f{\eb_{n-1}}{w_{n-1} x_{n-1}} \right),
\ee
\beq
\dot b_n = - \f{x_{n+1} b_{n+1}^2 - x_n b_n^2}{2 \gamma w_n x_n} -\f{\gamma}{2 x_n} 
+ \f{\gamma x_n}{2 (\eb_n)^2}.
\ee
In the equations of motion for $n=1$, there are no contributions from $n=0$, while the equations of motion for $n=0$ are trivial, $\dot E^b_0 = 0$ and $\dot b_0 = 0$, at least for the choice $\mH_0^{\rm disc}=0$.

Finally, the energy density is given by
\be \label{density-d}
\rho_n = - ~ \f{\mathcal{H}_n^{\rm disc}}{4 \pi w_n x_n \, |\eb_n|} - \f{1}{4 \pi w_n x_n |\eb_n|} \cdot \f{1}{G w_n} \left( \f{x_{n+1}^2}{|\eb_{n+1}|} - \f{x_n^2}{|\eb_n|} \right),
\ee
(the second term is the total derivative that was dropped in \eqref{eq:discHam} but is necessary to calculate $\rho_n$), while the total mass within a radius $x_n$ is
\beq
m_n = 4\pi \sum_{p=0}^n w_p x_p |E^b_p| \rho_p.
\ee

\section{Quantum theory}
\label{s.quantum}

The quantum theory for LTB space-times defined here is based upon the methods of loop quantum cosmology (LQC). Starting from the discretized classical theory given in Sec.~\ref{s.disc}, an LQC Hilbert space is defined at each point $x_n$ in the radial lattice; the total Hilbert space for LTB space-times is given by their tensor product. (Note that due to the gauge-fixing which gives a physical Hamiltonian, there is no distinction between a kinematical and physical Hilbert space in this case.) Then, the dynamics are generated by the Hamiltonian operator.

\subsection{Hilbert space}
\label{s.hil}

LQC techniques were first developed for cosmological space-times that are homogeneous and therefore have a finite number of degrees of freedom. This is not the case for LTB space-times, which allow fields that can depend on the radial coordinate. This adds a certain degree of complexity that is not present in the treatment of Friedman-Lema\^itre-Robertson-Walker (FLRW), Bianchi, or Kantowski-Sachs space-times.

To address this problem, we introduce an arbitrary discretization of the radial coordinate $x$ as explained in Sec.~\ref{s.disc}. At each node $n$, we can define a Hilbert space $\mathbb{H}_n$ following the standard methods of LQC, and then the full Hilbert space $\mathbb{H}$ for LTB space-times will be given by the tensor product $\mathbb{H} = \otimes \mathbb{H}_n$.

As usual in LQC, the elementary operators for each Hilbert space $\mathbb{H}_n$ correspond to densitized triads and complex exponentials of components of the Ashtekar-Barbero connection, from which it is possible to construct holonomies along coordinate edges.

To understand the relation between complex exponentials of $b$ and holonomies, recall that after the gauge-fixing imposed on the classical theory, it is the angular component of $A_a^i$ that has not been gauge-fixed and therefore we are interested in holonomies along paths tangent to, for example, $(\partial/\partial\theta)^a$ along the equator $\phi=\pi/2$; in general the paths of interest are portions of a great circle at different radii $x_n$. We use the `K' loop quantization \cite{Vandersloot:2006ws, Singh:2013ava} that is based on the `extrinsic curvature holonomy',
\begin{align} \label{holonomy}
h_{n,\theta}(\mu) &= \exp \left( \int_0^{\mu} \! \gamma K_\theta^i(x_n) \tau_i \, d\theta \right) \nn \\ &
= \exp \left( \int_0^{\mu} \! b_n \tau_2 \, d\theta \right)= \cos \left( \frac{\mu b_n}{2} \right) \mathbb{I} + 2 \sin \left( \frac{\mu b_n}{2} \right) \tau_2,
\end{align}
where $\mu$ is the coordinate length of the holonomy's path (in this case $\mu$ is the angle covered by the portion of the great circle). To be concrete, here we consider the great circle $\phi = \pi/2$, but the result would be similar for any other path along a portion of a great circle. The $\tau^i$ are a basis in the fundamental representation of the $\mathfrak{su}(2)$ Lie algebra with $\tau^i \tau^j = \frac{1}{2} \epsilon^{ij}{}_k \tau^k - \frac{1}{4} \delta^{ij} \mathbb{I}$, and $\mathbb{I}$ is the $2\times2$ identity matrix. Importantly, due to spherical symmetry the path integral along $\theta$ is trivial and easily evaluated---this would not be true for paths in the radial direction, but these are not needed in this case due to the gauge-fixing that has been imposed before quantization.

This defintion of the holonomy operator uses the fundamental representation of $\mathfrak{su}(2)$, also commonly utilized in LQC. Other choices are in principle possible and have been explored in LQC \cite{Vandersloot:2005kh, BenAchour:2016ajk, Amadei:2022zwp}. But the key physical feature of singularity avoidance that is central to our results is not affected by this choice: the curvature operator constructed from the holonomy is a bounded operator regardless of the representation of $\mathfrak{su}(2)$.

The non-trivial dependence on $b_n$ in \eqref{holonomy} is fully captured by the complex exponentials
\beq
\mN_n(\mu) = \exp(i \mu b_n)
\ee
of the component $b$ of the Ashtekar-Barbero connection. Given this result, we can choose the fundamental operators (at each node $n$) in the quantum theory to correspond to the densitized triad $E^b$ and complex exponentials of $b$.

It is convenient to express the Hilbert space $\mathbb{H}_n$ using the $E^b$-representation where the states $\ket{\eb}$ form an orthonormal basis 
\be
{}_{n}{\braket{\eb|\tilde{\eb}}_n} = \delta_{\eb\tilde{\eb}},
\ee
and are eigenkets of the triad operator
\be
\hat{E}^b_n \ket{\eb}_n = \eb \ket{\eb}_n.
\ee
As usual in LQC, the operators corresponding to $\mN_n(\mu)$ act as shift operators in this representation,
\be
\hat\mN_n(\mu) \ket{\eb}_n = \ket{\eb+\f{1}{w_n} \, \hbar G\gamma\mu}_n,
\ee
a direct calculation confirms that the commutator of these elementary operators matches the Poisson bracket of the equivalent observables on the classical phase space, up to the required factor of $i\hbar$,
\be
\{\mN_n(\mu),\eb_n\} = \f{iG\gamma\mu}{w_n}\mN_n(\mu)\qquad \rightarrow \qquad
[\hat\mN_n(\mu),\hat{E}^b_n] = -\f{\hbar G\gamma\mu}{w_n} \hat\mN_n(\mu).
\ee

Since $\hat{E^b_n}$ has a discrete spectrum, it is also necessary to define an inverse triad operator for each $\mathbb{H}_n$. There is considerable ambiguity in the possible choices here (see \cite{Singh:2013ava} for a discussion); we take the simplest choice possible \cite{Wilson-Ewing:2012dcf}:
\be \label{def-inv}
\widehat{\f{1}{E^b_n}} \ket{\eb}_n = 
\begin{cases} 
  0 & \text{if}\:\hat{E}^b_n\ket{\eb}_n=0, \\
  \left(E^b_n\right)^{-1}\ket{\eb}_n \quad & \text{otherwise}.
\end{cases}
\ee
In addition, note that $x_n^{-1}$ appears in the Hamiltonian, a term which could be problematic at the origin. Since these terms appear due to the gauge-fixing of $E^a=x^2$, there would normally be an inverse triad operator for $(E^a)^{-1}$ that vanishes for $E^a=0$, so we add the requirement that all terms in the Hamiltonian containing $x_n^{-1}$ vanish when $x_n = 0$.

Any state $\ket{\Psi}_n \in \mathbb{H}_n$ has the general form of a countable sum
\be
\ket{\Psi}_n = \sum_{\eb} \psi(\eb) \, \ket{\eb}_n,
\ee
with finite norm
\be
{}_n\braket{\Psi|\Psi}_n = \sum_{\eb} \left|\psi(\eb)\right|^2.
\ee

As already mentioned, the full Hilbert space $\mathbb{H}$ is given by the tensor product of the Hilbert spaces $\mathbb{H}_n$ located at each node, $\mathbb{H} = \otimes \mathbb{H}_n$, and the elementary operators $E^b_n, \mN_{n,\mu}$ acting at different nodes $n \neq m$ commute.

Finally, if the orientation of the densitized triad $E^b_i$ is flipped, $E^b_i(x) \to - E^b_i(x)$ (and also $b \to -b$), the space-time geometry remains the same. Therefore, we require that the parity transformation
\beq
\h\Pi \Psi(E^b_0, E^b_1, \ldots, E^b_n, \ldots) = \Psi(-E^b_0, -E^b_1, \ldots, -E^b_n, \ldots)
\ee
leaves the wave function invariant, namely
\beq
\h\Pi \Psi = \Psi.
\ee
As an aside, we note that it is possible to define a parity transformation at a given node which changes the sign of the $E^b_n$ argument of $\Psi$. However, this is not natural from a physical point of view since parity transformations are global operations and cannot act locally, therefore we do not introduce such an operator.

\subsection{Quantum dynamics}

In order to define an operator corresponding to the physical Hamiltonian in LQC, it is first necessary to construct a non-local operator corresponding to $\hat b_n$, which captures one of the components of the field strength through
\beq
\h F_{\theta\phi}{}^1(x_n) = \h b_n^2 \, \sin\theta.
\ee
Note that this operator is non-local since it is constructed from holonomies along edges of finite, not infinitesimal, length. Following the usual LQC procedure for the `K' quantization \cite{Vandersloot:2006ws, Singh:2013ava}, we use a holonomy of minimal physical length $\sim \lp$,
\beq \label{def-b-op}
\hat b_n = -2 \, {\rm Tr} \left( \f{\hat h_{n,\theta}(2\b\mu_n) - \hat h_{n,\theta}(-2\b\mu_n)}{2\b\mu_n} \tau_2 \right)
=\f{\sin \b\mu_n b_n}{\b\mu_n},
\ee
where $\b\mu_n$ is a coordinate length chosen so that the physical length of the path is $\sqrt\Delta$, where $\Delta \sim \lp^2$ is the smallest non-zero eigenvalue of the area operator in LQG. Of course, $\sin \b\mu_n b_n$ can be expressed in terms of $\h\mN_n(\b\mu)$ and $\h\mN_n(-\b\mu) = \h\mN_n(\b\mu)^\dag$.

As the holonomy \eqref{holonomy} follows a path where only the coordinate $\theta$ varies, it is straightforward to relate the coordinate length in the $\theta$ direction to the physical length through the metric \eqref{eq:metric-gf}, $|\dd s| = x |\dd\theta|$. Requiring that the physical length be given by $\sqrt\Delta$ sets
\beq
\b\mu_n = \f{\sqrt \Delta}{x_n},
\ee
which implies that the operator $\h b_n$ acts on basis states as
\beq
\h b_n \ket{E^b}_n = \f{x_n}{\sqrt\Delta} \Big( \, \ket{E^b + \f{1}{w_n x_n} \, G \hbar \gamma \sqrt\Delta} - \ket{E^b - \f{1}{w_n x_n} G \hbar \gamma \sqrt\Delta } \, \Big).
\ee
Interestingly, in this case $\b\mu_n$ is independent of phase-space variables, unlike in the LQC of homogeneous space-times. This is ultimately due to the gauge-fixing of $E^a = x^2$ in Sec.~\ref{s.ltb}.

With this definition, it is now possible to express the operator for the (discretized) physical Hamiltonian \eqref{eq:discHam} in terms of the elementary operators. As in the classical theory, the total physical Hamiltonian operator is given by a sum of operators, each corresponding to the physical Hamiltonian on one node of the lattice,
\beq
\h\mH = \sum_n \h\mH_n,
\ee
and using the elementary operators defined above, together with the definition for $\h b_n$ that captures the non-local curvature along angular directions, gives
\begin{align}
\label{eq:qHam}
\hat \mH_n = - \, \f{w_n}{2G\gamma} \Biggr[ 
& \frac{1}{\gamma x_n} \widehat{ \sqrt{|\eb_n|} } \f{1}{w_n} \left[ \frac{x_{n+1}^3}{\Delta} \, \sin^2 \left(\f{\sqrt\Delta b_{n+1}}{x_{n+1}} \right)-\frac{x_n^3}{\Delta} \, \sin^2 \left( \f{\sqrt\Delta b_n}{x_n} \right) \right] \widehat{ \sqrt{|\eb_n|} } \nn \\ 
& + \gamma x_n \abs{\widehat{ \f{1}{\eb_n}} } + \frac{\gamma}{x_n} |\hat{\eb_n}| \Biggr].
\end{align}
The only factor-ordering choice to be made is in the first term, where we chose a symmetric factor-ordering between $\hat E^b_n$ and $\hat b_n^2$.

Note that given the definition of the inverse triad operators and the requirement that terms in the Hamiltonian containing $x_n^{-1}$ also vanish for $n=0$ (due to inverse triad corrections for $E^a$ which has been gauge-fixed to $x_n^2$, see the discussion following the definition of the inverse triad operator in Sec.~\ref{s.hil}), it follows that $\hat \mH_0 = 0$. So although $\hat \mH_0$ is well-defined, it is trivial. For all other $n$, $\hat \mH_n$ is non-trivial and acts on $\mathbb{H}_n \otimes \mathbb{H}_{n+1}$, given the discretization scheme we chose for derivatives in Sec.~\ref{s.disc}. Assuming $E^b_n \neq 0$, the inverse triad operators give the second line in \eqref{def-inv} and the action of $\h\mH_n$ (for $n \ge 1$) on basis states $\ket{\eb}_n \otimes \ket{\eb}_{n+1} = \ket{\eb_n, \eb_{n+1}}$ is:
\begin{align}
\label{qHam-action}
\h\mH_n \ket{\eb_n, \eb_{n+1}} = 
& \, \f{x_{n+1}^3 \abs{E^b_n}}{8 G \gamma^2 \Delta x_n} \, \Big( \ket{\eb_n, \eb_{n+1} + \ell_{n+1}} + \ket{\eb_n, \eb_{n+1} - \ell_{n+1}} \Big) \nn \\
& - \f{x_n^2 \sqrt{|E^b_n|}}{8 G \gamma^2 \Delta} \, \Big( \sqrt{|\eb_n + \ell_n|} \, \ket{\eb_n + \ell_n, \eb_{n+1}} + \sqrt{|\eb_n - \ell_n|} \, \ket{\eb_n - \ell_n, \eb_{n+1}} \Big) \nn \\
&-\left[ \f{(x_{n+1}^3-x_n^3) \abs{E^b_n}}{4 G \gamma^2 \Delta x_n}
+ \frac{w_n \abs{\eb_n}}{2G x_n} + \frac{w_n x_n}{2G |\eb_n|} \right]
\ket{\eb_n, \eb_{n+1}},
\end{align}
where
\beq
\ell_n = \f{2 G \hbar \gamma \sqrt\Delta}{w_n x_n}.
\ee
Note that if $E^b_n = 0$, then all terms vanish exactly (the last term is zero due to the inverse triad operator).

Also, recall that $\h\mH$ is a true Hamiltonian (not a constraint) due to the dust-time gauge-fixing, and the Hamiltonian operator is related to the dust energy density operator,
\beq \label{rho-op}
\h\rho_n = - \f{1}{4 \pi w_n x_n} \left| \widehat{ \f{1}{\eb_n}} \right|^{1/2} \left[ \h\mH_n + \f{1}{G w_n} \left( x_{n+1}^2 \left| \widehat{ \f{1}{\eb_{n+1}}} \right| - x_n^2 \left| \widehat{ \f{1}{\eb_n}} \right| \right) \right] \left| \widehat{ \f{1}{\eb_n}} \right|^{1/2} ,
\ee
while the mass operator is given by
\beq
\h m_n = 4 \pi \sum_{p=0}^n w_p x_p \sqrt{\abs{\hat E^b_p}} \hat \rho_p \sqrt{\abs{\hat E^b_p}},
\ee
where we have again used a symmetric factor-ordering for both operators.

Finally, to reconstruct all of the components in the metric \eqref{eq:metric-gf}, it is also necessary to define an operator corresponding to $N^x$. Recall that the flat FLRW space-time (minimally coupled to dust) is a particular solution of the LTB space-time; starting from the standard flat FLRW line element in spherical coordinates $\dd s^2 = -\dd t^2 + a(t)^2 (\dd r^2 + r^2 \dd\Omega^2)$ where $a(t)$ is the scale factor, the coordinate transformation $x = a(t) \cdot r$ gives a line element that has precisely the form \eqref{eq:metric-gf} with $E^b = x$ and $N^x = -xH$, where $H$ is the Hubble rate. In LQC the Hubble rate in terms of the phase space variables is $H = (\gamma \sqrt\Delta)^{-1} \sin (\b\mu b) \cos(\b\mu b)$, see, e.g., \cite{Rovelli:2013zaa} (where we have simplified the notation by removing hats, and denoted the non-zero component of the Ashtekar-Barbero connection for the flat FLRW space-time by $b$). This suggests that the appropriate operator for $N^x$ in LTB space-times is
\beq \label{def-nx}
\h N^x_n = -\f{x_n}{\gamma \sqrt\Delta} \sin \left( \f{\sqrt\Delta b_n}{x_n} \right) \cos \left( \f{\sqrt\Delta b_n}{x_n} \right). 
\ee
This choice has the correct classical limit, and it also agrees with earlier work \cite{Gambini:2020nsf, Kelly:2020uwj} (although it differs from the choice proposed in \cite{Gambini:2020qhx}). This operator is not obtained by simply replacing the classical $b_n$ in $N^x$ by the operator $\h b_n$ defined in \eqref{def-b-op}; nonetheless, this is a natural choice for the operator as it ensures that the homogeneous sector of (marginally-trapped) LTB space-times agrees with earlier results obtained for flat FLRW space-times in LQC. (Note that there is a priori no guarantee that $(N^x)^2 \sim b^2$ and the field strength $F_{\theta\phi}{}^1 \sim b^2$ will be represented by the same operator in the quantum theory; it is possible for $\hbar$ corrections to modify these terms differently.)

As an aside, we mention two points concerning the operator $\h N^x_n$. First, even though as argued above we find \eqref{def-nx} to be the most strongly motivated definition for $\h N^x_n$, it is nonetheless possible to make other choices, see, e.g., \cite{Gambini:2020qhx} for an alternate choice in vacuum space-times. A different choice affects quantitative results, but our numerical evidence suggests that the key qualitative properties of the (effective) LQC dynamics for LTB space-times, such as the non-singular bounce, shock wave formation, and black hole lifetime are not significantly modified by different choices for $N^x$---this is due in large part to fact that the evolution of $b$ and $\eb$ does not depend on the reconstruction of the metric via \eqref{def-nx}. A second technical point is the following. Although this has so far not been done, it should be possible to define a consistent quantum theory for LTB space-times without first imposing any gauges before quantization, and instead impose appropriate gauges only after the theory has been quantized. Gauge fixing after quantization would require introducing operators corresponding to gauge-fixing conditions that ensure the gauge choices are preserved dynamically, and these gauge-fixing conditions will fix the Lagrange multipliers, here corresponding to the lapse and shift. The classical gauge-fixing conditions for the dust-time and areal gauges in LTB space-times are known, and it would be interesting to determine their form as operators in LQC; in particular, it will be necessary to express the components of the Ashtekar-Barbero connection in terms of holonomies. Although deriving the specific form of such gauge-fixing conditions for LQC lies beyond the scope of this paper, it can be shown that since the operator for $N^x$ is not related to the classical one by a simple `polymerization' where each $b_n$ term is replaced by $\hat b_n$, similarly the LQC gauge-fixing conditions cannot be obtained simply by taking the classical expressions and performing a direct polymerization \cite{Giesel:2021rky}. This is not surprising for two reasons: first, the LQC shift vector is not given by the direct polymerization of the classical expression, and second, the non-gauge-fixed LQC scalar and diffeomorphism constraints for the LTB space-times are not related to the classical expressions through a direct polymerization, since this would result in a quantum constraint algebra which does not close and hence would give an inconsistent theory. A more careful and sophisticated treatment is needed to construct the non-gauge-fixed LQC theory for LTB space-times; this presumably requires following LQG more closely and avoiding shortcuts like polymerization which appear to work only in the simplest contexts.

This completes the definition of the loop quantization of LTB space-times. We next derive effective equations which capture important quantum gravity effects, and are significantly easier to solve than the full quantum dynamics.

\section{Effective Theory} 
\label{s.effective}

Loop quantum cosmology effective equations are functions on the classical phase space that include some quantum corrections in the form of terms containing $\hbar$ that arise from, e.g., holonomy corrections. It would of course be preferable to solve the full quantum dynamics, but this is a technically challenging problem---deriving semiclassical physics would require constructing the unitary evolution operator for the Hamiltonian and calculating its action on semiclassical states. Instead, as a first step, we extract and solve the effective equations with holonomy corrections from the LTB quantum theory defined in Sec.~\ref{s.quantum}.

For homogeneous cosmological space-times, it has been shown that the quantum dynamics of sharply-peaked states is well-approximated by a set of effective equations, and that these effective equations remain a good approximation to the quantum dynamics for sharply peaked states even when the space-time curvature becomes Planckian, so long as the relevant length scales remain large compared to $\lp$ \cite{Ashtekar:2006wn, Taveras:2008ke, Rovelli:2013zaa, Bojowald:2015fla}. The main approximation underlying the effective equations is that they assume quantum fluctuations are negligible, but include important effects like the fundamental quantum discreteness encoded by the non-local curvature operator in the Hamiltonian.

In addition to results in cosmology, some early work on black holes suggests that effective dynamics are also reliable for black hole space-times with a large mass, at least for (i) states that are sharply peaked and (ii) so long as one only probes length scales $\ell \gg \lp$ \cite{Zhang:2021xoa}. If these two conditions are satisfied, then the effective dynamics are expected to provide a good approximation to the full quantum dynamics, even when the curvature is Planckian.

In this section, we define the LQC effective Hamiltonian for LTB space-times in the dust-time and areal gauges, derive the LQC effective dynamics, and find analytic weak solutions for some simple configurations.

\subsection{Effective Dynamics}

The effective equations for LTB space-times can be derived from the quantum theory by expressing the Hamiltonian operator \eqref{eq:qHam} as a function on the classical phase space. For the LTB space-time, this gives an effective Hamiltonian (for the discretized theory) composed of the sum over $n$ of
\begin{align}
\mH_n^{\rm eff} = - \, \f{1}{2G\gamma} \Biggr[ 
& \frac{\eb_n}{\gamma w_n x_n} \left( \frac{x_{n+1}^3}{\Delta} \, \sin^2 \f{\sqrt\Delta b_{n+1}}{x_{n+1}} - \frac{x_n^3}{\Delta} \, \sin^2 \f{\sqrt\Delta b_n}{x_n} \right) 
 + \f{\gamma x_n}{\eb_n} + \frac{\gamma \eb_n}{x_n} \Biggr],
\end{align}
assuming $\eb>0$ for the sake of simplicity. The continuum limit is easily taken; the resulting Hamiltonian is
\beq
\int \dd x \, \mH_{\rm phys}^{\rm eff} = - \f{1}{2G\gamma} \int \dd x \left[ \f{\eb}{\gamma \Delta x} \pd_x \left( x^3 \sin^2 \f{\sqrt\Delta b}{x} \right)
+ \f{\gamma x}{\eb} + \frac{\gamma \eb}{x} \right].
\ee
The transition from the Hamiltonian operator to this effective expression requires taking the expectation of the former in a semiclassical Gaussian state peaked on a phase space point (see, e.g., \cite{Husain:2006uh}), and then taking the continuum limit. This is the same process used to obtain the effective LQC dynamics, where the Hamiltonian operator with matter is also unbounded. A shortcut that is known to work in LQC, and one we use here, is to replace the quantum operators with the appropriate phase space functions. Note that the effective dynamics do not take into account dispersion effects of the Hamiltonian operator captured by expressions as such $\Delta (\mH_n)^2\equiv\langle \hat \mH_n^2 \rangle - \langle \hat \mH_n \rangle^2$ for expectation values in semiclassical states, which are of higher order in $\hbar$.

The Poisson bracket is the same as in the classical theory,
\be
\{b(x_1), E^b(x_2)\} = G \ga \, \de(x_1 - x_2),
\ee
and the effective equations of motion are generated, as usual, by taking the Poisson brackets of the dynamical variables with the effective Hamiltonian, $\int \! \dd x \, \mH_{\rm phys}^{\rm eff}$, giving
\bea
\dot b &=& \f{\ga x}{2 (E^b)^2} - \f{\ga}{2x} - \f{x}{\ga \De} \sin \f{\sqrt\De}{x} b \left[ \f{3}{2} \sin \f{\sqrt\De}{x} b + x \, \partial_x \sin \f{\sqrt\De}{x} b \right], \\
\dot E^b &=& - \, \f{x^2}{\ga \sqrt \De} \, \partial_x \left(\f{E_b}{x}\right) \sin \f{\sqrt\De}{x} b ~ \cos \f{\sqrt\De}{x} b.
\eea

As in the classical theory, $b$ and $\eb$ determine the metric,
\be \label{modmetric}
\dd s^2 = - \dd t^2 + \f{E_b^2}{x^2} (\dd x + N^x \dd t)^2 + x^2 \dd\Omega^2,
\ee
where the shift vector is related to $b$ by
\be
N^x = - \, \f{x}{\ga \sqrt\De} \sin \f{\sqrt\De}{x} b ~ \cos \f{\sqrt\De}{x} b,
\ee
taking the prescription \eqref{def-nx} used for the quantum theory.

Finally, the energy density in the effective theory follows from \eqref{rho-op},
\be
\rho = \f{1}{8 \pi G \gamma \, x \eb} \left[ \f{\eb}{\gamma \Delta x} \pd_x \left( x^3 \sin^2 \f{\sqrt\Delta b}{x} \right)
+ \f{\gamma x}{\eb} + \frac{\gamma \eb}{x}
- 2 \gamma \, \pd_x \left( \f{x^2}{\eb} \right) \right].
\ee
We note that the area gap $\Delta$ and the Barbero-Immirzi parameter $\gamma$ often combine in the form $\gamma^2 \Delta$, and many observables of physical interest depend only on this combination. This also occurs in LQC and seems to be a general feature of loop quantized symmetry-reduced systems. In contrast, in LQG the quantities $\gamma$ and $\Delta$ are distinct (although related, since $\Delta$ is proportional to $\gamma$), and can be distinguished for example by measuring the spectrum of the LQG area operator.

\subsection{Marginally-trapped solutions}

As in the classical theory, $\eb = x$ corresponds to the effective version of the marginally-trapped solutions. For this family of solutions, $b$ is the only dynamical degree of freedom left, satisfying the equation of motion
\be \label{eomb}
\dot b + \f{1}{2 \ga\De x} \pd_x \left( x^3 \sin^2 \f{\sqrt\De \, b}{x} \right) = 0.
\ee
The energy density is closely related to $b$ through
\beq \label{q-density}
\rho = \f{1}{8 \pi G \ga^2 \De \, x^2} \, \pd_x \left( x^3 \sin^2 \f{\sqrt\De \, b}{x} \right),
\ee
which can also be expressed in terms of $\dot b$, although this relation will not be necessary for us here. Expanding the derivative,
\beq
\rho = \f{1}{8 \pi G \ga^2 \De} \left (3 \sin^2 \f{\sqrt\De \, b}{x} + 2 \sqrt\Delta \sin \f{\sqrt\De \, b}{x} \cos \f{\sqrt\De \, b}{x} \, \Big{(}\pd_x b-\f{b}{x}\Big{)} \right).
\ee
In the homogeneous limit of $\pd_x b = 0$, the energy density is bounded above by $\rho_c = 3 / 8 \pi G \ga^2 \De,$ the critical energy density for spatially flat FLRW space-times in LQC.

It is simpler and more intuitive to give initial conditions in terms of the density $\rho(x, t_0)$; these can be translated into initial conditions for $b$ through
\beq
\label{eq:initb}
b(x, t_0) = - \f{x}{\sqrt\De} \arcsin \left( \sqrt{ \f{8 \pi G \ga^2 \De}{x^3} \int_0^x \dd \t x ~ \t x^2 \rho(\t x,t_0) } \right) ~.
\ee
When taking the square root, we chose an overall negative sign; this choice means that the initial data corresponds to a collapse scenario. On the other hand, taking the positive root would correspond to an LTB space-time with dust initially moving outwards.

For collapse models, an important quantity is the outgoing null expansion $\theta_+$, which is used to locate marginal apparent horizons via $\theta_+ = 0$. For the LTB metric \eqref{modmetric}, with $\eb=x$, the outgoing null expansion is 
\beq
\theta_+ = \f{2}{x} (1 - N^x) = \f{2}{x} + \f{1}{\ga \sqrt\De} \sin \f{2 \sqrt\De \, b}{x}.
\ee
Similarly, the in-going null expansion is $\theta_- = -(2/x) \cdot (1 + N^x)$.

It is possible to find implicit solutions for $b$ by using the method of characteristics, namely by finding curves in the $(x,t)$ plane parametrized by $\lambda$ such that along these curves the partial differential equation \eqref{eomb} becomes a set of coupled ordinary differential equations. Solving the set of coupled ordinary differential equations along each of these curves gives the full solution, although in implicit form in terms of the parameter $\lambda$.

It is convenient to first introduce the variable $\beta = \sqrt\De \, b / x$, in which case the equation of motion becomes
\beq \label{eom-beta}
\dot \beta + \f{1}{2 \ga \sqrt\De x^2} \pd_x \left( x^3 \sin^2 \beta \right) = 0.
\ee
As an aside, note that this equation can be rewritten as
\beq \label{eom-beta-rho}
\dot \beta = - 4 \pi G \ga \sqrt\De \, \rho,
\ee
and since $\rho \ge 0$, it follows that $\beta$ is monotonically decreasing.

Taking a parametrized curve $(x(\lambda), t(\lambda))$ and evaluating $\beta(\lambda)$ at points along this curve gives
\be
\f{\dd\beta}{\dd\lambda} = \f{\del\beta}{\del t} \f{\dd t}{\dd\lambda} + \f{\del\beta}{\del x} \f{\dd x}{\dd \lambda}.
\ee
By choosing a curve such that
\be
\f{\dd\beta}{\dd\lambda} = -\f{3}{2} \sin^2\beta,
\ee
the equation of motion for $\beta$ implies that
\be
\f{\dd x}{\dd\lambda} = x \sin\beta \cos\beta, \qquad \f{\dd t}{\dd\lambda} = \gamma\sqrt\Delta.
\ee
Solving these ODEs first gives $\lambda = t/\gamma\sqrt\Delta$ (setting $\lambda(t=0) = 0$), and
\be
\label{eq:effchar}
\cot \big(\beta(\lambda)\big) = \cot \beta_0 + \f{3\lambda}{2}, \qquad
\left( \f{x}{x_0} \right)^3 = \f{\sin^2 \beta_0}{\sin^2 \beta}.
\ee
For each curve, the initial conditions are $t(\lambda=0) = 0$ and $x(\lambda=0) = x_0$. For such a curve, $\beta(\lambda=0) = \beta_0(x_0)$, and we have the exact implicit solutions for $\beta(x,t)$. These can be inverted numerically everywhere, except where characteristics cross---at these points the method of characteristics fails, and it is necessary to search for weak solutions to the equation of motion. As mentioned previously, it is always possible in the classical theory to restrict initial data such that caustics do not form \cite{Hellaby:1985zz, Nolan:2003wp, Booth:2005ng}, but as we explain next this is not possible in the effective quantum theory due to the singularity avoidance and bounce: shock waves are a general feature of quantum dust collapse.

As in the classical case, shock formation occurs when the Jacobian for the transformation $(\lambda,x_{0})\rightarrow (t,x)$ vanishes. Since $\partial_\lambda t = \gamma \sqrt\Delta$ and $\partial_{x_0} t = 0$, the Jacobian vanishes if and only if $\partial_{x_0} x = 0$. Rewriting
\beq 
x = \left( x_0^3 \sin^2 \beta_0 \Big[1 + (\cot \beta_0 + \tfrac{3}{2}\lambda)^2\Big] \right)^{1/3},
\ee
a numerical investigation finds that the derivative
\be
\pd_{x_0}x = \f{1}{3} \left(\f{\sin^2 \beta_0}{\sin^2 \beta}\right)^{1/3} \left[ 3 + 2 x_0 \beta_0' \left( \cot \beta_0 - \frac{1}{\sin^2 \beta_0} \cdot \frac{\tfrac{3}{2}\lambda + \cot \beta_0}{1+\left(\tfrac{3}{2}\lambda + \cot \beta_0 \right)^2} \right)\right]
\ee
generically vanishes for initial data with in-falling dust that satisfies the two following properties: the initial density $\rho_0$ is not zero everywhere, and there is an exterior vacuum region $x \ge x_e$ where $\rho_0=0$. Physically, this can be understood by following characteristics in the interior and exterior regions. In the interior, the characteristics for dust particles will eventually bounce due to LQC effects and move outwards, while in the vacuum exterior region characteristic curves will always move inwards. As a result, these two families of characteristic curves must eventually cross; see also the discussion in \cite{Schmitz:2019jct}. (Note that characteristic curves where $\rho \neq 0$ may also cross, but this depends quite sensitively on the initial configuration of $\rho_0$ and will not always occur, while characteristic curves for the interior matter region and the exterior vacuum region will cross after the bounce.) Therefore, for this large class of initial data, characteristics will cross, showing that it is necessary to allow for weak solutions to the effective dynamics.

\subsection{Weak solutions}

As characteristic curves will cross, weak solutions to the dynamics must be considered. As reviewed in Sec.~\ref{s.weak}, it is helpful to write the dynamics in the form of a conservation law $\dot u + \pd_x[f(u,x)] = 0$, which can in turn be expressed as an integral equation. Then, the speed of a shock wave located at $x=L(t)$ is given by $\dd L / \dd t = [f(u,x)]/[u]$, where $[g(x)] = \lim_{x\to L^+} g(x) - \lim_{x\to L^-} g(x)$.

To rewrite the equation of motion \eqref{eomb} as a conservation law, it is useful to introduce the variable $B = \sqrt\Delta \, x \, b$, whose dynamics are given by
\beq
\dot B + \pd_x \left( \f{x^3}{2 \ga \sqrt\De} \sin^2 \f{B}{x^2} \right) = 0.
\ee
Then, the speed of any shock waves that may form is given by
\beq \label{sw-beta}
\f{\dd L}{\dd t} = \f{L^3}{2 \ga \sqrt\De} \cdot \f{[\sin^2 \f{B}{x^2}]}{[B]} = \f{L}{2 \ga \sqrt\De} \cdot \f{[\sin^2 \beta]}{[\beta]}.
\ee
Here we give the expression in terms of both of the variables $B$ and $\beta = B/x^2$; 
depending on the calculation, one variable may be more convenient than the other.

The Oppenheimer-Snyder model and the thin shell solution can be solved analytically; the general case for arbitrary initial data profiles requires a numerical solution.

\subsubsection{Oppenheimer-Snyder solution}
\label{s.os-collapse}

The Oppenheimer-Snyder collapse model has two regions: a `star' interior region $x < L(t)$ where the energy density $\rho$ is radially constant (but grows with time as the star collapses and becomes denser), and a vacuum exterior region $x > L(t)$ where $\rho = 0$. We will denote the interior region with the index $i$ and the exterior region with the index $e$.

The solution for the $\rho_e = 0$ vacuum exterior region follows from \eqref{q-density},
\beq
\sin \beta_e = - \sqrt \f{\ga^2\De R_S}{x^3},
\ee
where the minus sign is due to the attractive nature of the gravitational field generated by the dust field in the interior, and the constant of integration is chosen to match the classical solution. As usual, $R_S = 2GM$ is the classical Schwarzschild radius and $M$ is the total mass of the star (which is a constant of the motion). Also, note that $\dot \beta_e = 0$.

Similarly, for the interior region \eqref{q-density} shows that if $\pd_x\rho = 0$, then $\pd_x \beta = 0$ also, and \eqref{eom-beta} becomes
\beq
\dot\beta_i = - \f{3 \sin^2 \beta_i}{2 \ga \sqrt\De},
\ee
with the solution
\beq
-\cot \beta_i = \f{3 (t - t_0)}{2 \ga \sqrt\De} \qquad \Rightarrow \qquad \sin \beta_i = \f{-1}{\sqrt{1 + 9t^2/4\ga^2\De}}.
\ee
The minus sign indicates that the Oppenheimer-Snyder star is collapsing, rather than expanding, and in the second relation we have fixed the constant of integration $t_0 = 0$. Since $\beta_i$ is monotonically decreasing, as seen in \eqref{eom-beta-rho}, it follows that for $t < 0$, then $-\pi/2 < \beta_i < 0$, while $-\pi < \beta_i < -\pi/2$ for $t>0$. As a result, $\beta_i$ itself is given by
\beq
\beta_i =
\begin{cases} - \arcsin \f{1}{\sqrt{1 + 9t^2/4\ga^2\De}}, \qquad & {\rm for~} t < 0, \\
-\pi + \arcsin \f{1}{\sqrt{1 + 9t^2/4\ga^2\De}}, \qquad & {\rm for~} t > 0, \end{cases}
\ee
keeping in mind that $-\pi/2 < \arcsin x < \pi / 2$.

In terms of the gauge-fixed metric \eqref{modmetric}, the interior is a flat FLRW space-time with a dust field whose energy density evolves as
\beq \label{rho-os}
\rho_i = \f{3}{2 \pi G \left( 9t^2 + 4\ga^2\De \right) }.
\ee

During the $t < 0$ contracting phase of the star, $\beta$ is continuous across the boundary between the interior and exterior solutions (although not differentiable), so there is no shock wave and the location of the boundary $x=L(t)$ can be determined by requiring that $\beta$ be continuous across the boundary, giving \cite{Kelly:2020lec}
\beq \label{L-cont}
L = \left( \f{9 R_S t^2}{4} + \ga^2\De R_S \right)^{1/3}.
\ee
Using this expression to rewrite the energy density of the interior, it is clear that $L(t)$ plays the role of the scale factor for the FLRW-like interior, with
\beq
\rho_i = \f{3 M}{4 \pi L^3},
\ee
and the effective Friedman equation that $L$ satisfies during the collapse phase is
\beq \label{lqc-fried}
\left( \f{\dot L}{L} \right)^2 = \f{8 \pi G}{3} \rho \left(1 - \f{\rho}{\rho_c}\right),
\ee
which is exactly identical to the LQC effective Friedman equation for flat FLRW space-times.

A bounce occurs at $t=0$ when $\beta_i = - \pi/2$. The bounce occurs both in the radius $L$ of the collapsing star, which reaches a minimum and begins to increase, and also in the energy density $\rho$, which reaches a maximum precisely equal to $\rho_c = 3/8\pi G \ga^2\De$, the critical energy density in LQC, and then decreases after the bounce.

The post-bounce dynamics are significantly different from the collapse, because after the bounce $\beta_i < - \pi / 2$ while $\beta_e > -\pi /2$, showing that a discontinuity in $\beta$ has formed: there is now a shock in the gravitational field. Since there is a shock, $L$ no longer follows the dynamics given by \eqref{lqc-fried} which holds only when the solution is continuous; instead the motion of the front of the shock wave during the expanding phase is determined by the Rankine-Hugoniot condition \eqref{sw-beta},
\beq \label{os-shock}
\f{\dd L}{\dd t} = \f{\gamma\sqrt\Delta / 2}{\pi - \arcsin \sqrt \f{\gamma^2\Delta R_S}{L^3} - \arcsin \sqrt \f{\gamma^2\Delta}{\tfrac{9}{4}t^2 + \gamma^2\Delta}} \cdot \left( \f{R_S}{L^2} - \f{L}{\tfrac{9}{4} t^2 + \gamma^2\Delta} \right).
\ee

In the interior, away from the shock, the energy density after the bounce decreases following \eqref{rho-os}, but $L$ moves more slowly after the bounce, as compared to before, and the outside region remains vacuum. The combination of these effects implies that a growing fraction of the dust field becomes trapped at the boundary $x=L(t)$ in the expanding phase,
\beq
\rho_{\rm int} = \f{3}{2 \pi G \left( 9t^2 + 4\ga^2\De \right) } \Theta\Big(L(t) - x\Big) + \left( M - \f{2 L^3}{G \left( 9t^2 + 4\ga^2\De \right) } \right) \delta\Big(L(t) - x\Big),
\ee
where $\Theta(x)$ is the Heaviside function.

Since the energy density of dust in the interior quickly decays, soon after the bounce it is reasonable to approximate the interior by a flat Minkowski space-time where $\beta = -\pi$ and $\sin\beta = 0$. Further, soon after the bounce $L \gg L_{\rm bounce} = (\ga^2 \Delta R_S)^{1/3}$ and it is possible to neglect terms of the order $L_{\rm bounce}/L$. With these two approximations, the speed of the shock \eqref{os-shock} reduces to
\beq
\f{\dd L}{\dd t} \approx \f{\gamma\sqrt\Delta R_S}{2 \pi L^2},
\ee
which can be solved to give
\beq \label{sol-shock-os}
L(t) \approx \left( \f{3 \ga \sqrt\Delta R_S \, t}{2 \pi} \right)^{1/3},
\ee
although this result is a good approximation only for $t \gg t_{\rm Pl}$. At the bounce $L(0)$ does not vanish, rather $L(0) = L_{\rm bounce} = \ga^2 \Delta R_S$.

With these results, it is straightforward to determine the space-time metric; putting $E^b=x$ into \eqref{modmetric},
\beq
\dd s^2 = -\dd t^2 + \Big( \dd x + N^x \dd t \Big)^2 + x^2 \dd\Omega^2,
\ee
the lapse $N^x = - (x / \gamma \sqrt\Delta) \cdot \sin\beta \cos\beta$ is \cite{Kelly:2020lec}
\be \label{eq:confshift}
N^x = 
\left\{
\begin{aligned}
& - \f{6 x t}{9 t^2 + 4 \gamma^2 \Delta} \quad & \mbox{if } x \le L(t), \\
& \sqrt{ \f{R_S}{x} \left( 1 - \f{\gamma^2 \Delta R_S}{x^3} \right)} & \mbox{if } x > L(t);
\end{aligned}
\right.
\ee
where $L(t)$ is given by \eqref{L-cont} for the collapse $t < 0$ portion of the space-time, while $L$ expands more slowly following \eqref{os-shock} for $t > 0$ after the bounce. For more details on the exterior vacuum solution see \cite{Kelly:2020uwj}, and for the interior solution see \cite{Kelly:2020lec}.

In addition, it is possible to calculate the lifetime of an Oppenheimer-Snyder black hole solution, defined as the (proper) time interval between the formation of the outer apparent horizon, and its eventual disappearance when the outgoing shock wave reaches it, as measured by a distant observer. Assuming the distant observer detects lightlike signals emitted from the surface $L$ just before the formation of the black hole and just after the emergence of the shock from the outer apparent horizon, this observer's proper time interval is simply given by the coordinate interval $T = t_2 - t_1$ between these two events \cite{Kelly:2020lec}. For this calculation, we make the approximation that the apparent horizon is located at $R_S$, neglecting corrections to the location of the horizon of the order $\sim \lp^2/R_S$ that are negligible for black holes with $M \gg m_{\rm Pl}$.

The black hole forms at $t=t_1$ when $L=R_S$ during the collapse phase, so by inverting \eqref{L-cont} we find
\beq
t_1 = - \f{2}{3} \sqrt{R_S^2 - \ga^2 \De} \, \approx \, - \, \f{2 R_S}{3},
\ee
showing that the collapse time of the Oppenheimer-Snyder star is of the order of $\sim M$.

On the other hand, the apparent horizon vanishes once the outgoing shock wave reaches $L = R_S$, and the solution \eqref{sol-shock-os} gives
\beq \label{est-t2}
t_2 \approx \f{2 \pi R_S^2}{3 \ga \sqrt\De}.
\ee
As an aside, we mention that it is also possible to numerically integrate the exact equation
\beq
t_2 = \int_{L_{\rm bounce}}^{R_S} \left(\f{\dd L}{\dd t} \right)^{-1} \dd L,
\ee
with $\dd L / \dd t$ given by \eqref{os-shock}, and the result of this calculation is in excellent agreement with the approximate solution \eqref{est-t2}. This shows that the time between the bounce and the disappearance of the outer apparent horizon scales as $\sim M^2 / m_{\rm Pl}$, in agreement with an earlier estimate \cite{Kelly:2020lec}.

Combining these two results and keeping only the dominant term (assuming $M \gg m_{\rm Pl}$), the lifetime of an Oppenheimer-Snyder black hole is predicted to be
\beq \label{life-os}
T \approx \f{2 \pi R_S^2}{3 \ga \sqrt\De}.
\ee
As we shall discuss in more detail in Sec.~\ref{s.impl}, this prediction for the lifetime of a black hole is shorter than the Page time and has important implications for the information loss problem.

\subsubsection{Thin shell solution}

Another interesting solution to consider is a thin shell, where the interior is Minkowski with $\sin \beta_i = 0$, and the exterior is Schwarzschild with $\sin \beta_e = - \sqrt{\ga^2\De R_S / x^3}$~; these two solutions are separated by a thin shell of total mass $M$ located at $x=L(t)$,
\beq
\rho = \f{M}{4 \pi x^2} ~ \delta \Big( x - L(t) \Big).
\ee
There are two cases of interest, the contracting case when $\beta_i = 0$, and the expanding case $\beta_i = -\pi$.

Before studying the dynamics, it is important to note that although there is a bounce, a contracting thin shell solution does not immediately become an expanding thin shell solution when the bounce occurs. The key point is that a thin shell of mass $M$ can reach a minimal radius $L_{\rm min} = (\ga^2 \De R_S)^{1/3}$. When the shell reaches this minimal radius, the thin shell splits, with a portion continuing to fall inwards and another portion being scattered outwards. The scattering process is continuous, with a broad distribution of dust being scattered backwards until there is a bounce at the origin (the total mass remains $M$ throughout). This process can be seen in more detail using the numerical methods described in Sec.~\ref{s.numerics}. For this reason, the ingoing and outgoing shell solutions must be treated separately. (Another way to see that the ingoing thin shell does not immediately become an outgoing shell after reaching $L_{\rm min}$ is that an ingoing shell has $\beta_i = 0$, while an outgoing shell has $\beta_i = -\pi$, so the first cannot just bounce at $L_{\rm min}$ and become the second without any changes to the interior.)

For the case of an ingoing thin shell, the shock speed relation \eqref{sw-beta} gives
\beq
\f{\dd L}{\dd t} =
- \, \f{2 \ga \sqrt\De R_S}{L^2 \arcsin \sqrt{\ga^2 \De R_S / L^3}},
\ee
this can be integrated to give an analytic implicit solution for $L(t)$, although it is not especially transparent or useful. Instead, the collapse time $t_{\rm in}$ can be obtained by integrating ($\dd L / \dd t)^{-1}$ from $L=R_S$ (again neglecting small Planckian corrections to the location of the outer apparent horizon) to $L_{\rm min} = (\ga^2 \De R_S)^{1/3}$; the result is
\beq
t_{\rm in} = - \, \int_{R_S}^{L_{\rm min}} \!\!\! \f{2 L^2}{\ga \sqrt\De R_S} \arcsin \sqrt \f{\ga^2 \De R_S}{L^3} ~ \dd L \, \approx \, \f{2 R_S}{3},
\ee
dropping subleading corrections of order $\hbar$. As expected, the time for collapse is $t_{\rm in} \sim M$, as is the case for the Oppenheimer-Snyder collapse (in this particular case, even the prefactors of the leading order term match for the thin shell and Oppenheimer-Snyder solutions).

Switching now to the outgoing case, the shock speed \eqref{sw-beta} becomes
\beq
\label{eq:outshell}
\f{\dd L}{\dd t} =
\f{\ga \sqrt\De R_S}{2 L^2 ( \pi - \arcsin \sqrt{\ga^2 \De R_S / L^3})},
\ee
so the time for the shell to travel from $L_{\rm min}$ to $R_S$ is
\beq
t_{\rm out} = \int^{R_S}_{L_{\rm min}} \f{2 L^2}{\ga \sqrt\De R_S} \left( \pi - \arcsin \sqrt \f{\ga^2 \De R_S}{L^3} \right) ~ \dd L \, \approx \, \f{2 \pi R_S^2}{3 \ga \sqrt\De},
\ee
again only keeping the leading order term when evaluating the integral.

In general, the lifetime of a black hole can be split into three parts: a collapse time, a bounce time, and an outgoing time. (The bounce time may be zero in some cases where the bounce occurs simultaneously everywhere, like in the Oppenheimer-Snyder model, but this will not necessarily always be true.) The lifetime of a black hole will be dominated by the outgoing time $t_{\rm out}$,
\beq \label{life-ts}
T \approx t_{\rm out} \approx \f{2 \pi R_S^2}{3 \ga \sqrt\De}.
\ee
Note that the leading order contribution to $T$ is identical for the thin shell and Oppenheimer-Snyder solutions. This is not surprising because after the bounce, the dust energy density $\rho$ in the Oppenheimer-Snyder interior rapidly decays to the point where the interior is well approximated by Minkowski space and all of the matter is located at the shock. In other words, soon after the bounce the Oppenheimer-Snyder solution becomes (to an excellent approximation) an outgoing thin shell. Further, as shall be seen in Sec.~\ref{s.numerics}, numerics show that this occurs quite generally: for a large class of initial density profiles for the collapse, after the bounce the outgoing shock wave rapidly tends to the outgoing thin shell solution. For this reason, the lifetime of a black hole appears to be universal to leading order with $T = 2 \pi R_S^2 / (3 \ga \sqrt\De) + {\cal O}(M)$.

\subsubsection{Conformal diagram}
\label{s.conf}

The conformal diagram for the effective vacuum solution has already been studied in considerable detail \cite{Munch:2021oqn}, but there are some important differences in the conformal diagram once matter (in this case dust) is included. For concreteness, we will sketch the conformal diagram for the Oppenheimer-Snyder collapse model derived in Sec.~\ref{s.os-collapse}, but we expect a qualitatively similar diagram for most of the main features for other solutions to the LQC effective dynamics for LTB space-times that start from a collapse that leads to the formation of a black hole---this expectation is met for the numerical solutions we obtain in Sec.~\ref{s.numerics}.

In the Oppenheimer-Snyder model, during collapse the radius $L(t)$ of the dust sphere is given by \eqref{L-cont}, while after the bounce the shock wave moves outwards following \eqref{os-shock}, whose solution is approximated by \eqref{sol-shock-os} for times $t \gg t_{\rm Pl}$.

In broad strokes, the conformal diagram shows the following events and processes. First, a pair of apparent horizons appears when $L = R_S$ during the collapse, the outer horizon is null and remains at the radius $x=R_S$, while the inner horizon lies inside $L$ for almost all of the collapse. Second, the inner horizon crosses outside $L$ only a short (Planckian) time before the bounce, at this point the inner horizon becomes null and remains at the same radius $x_{\rm inner}$ until the fourth stage. Third, after the inner horizon stops at $x_{\rm inner}$ the entire spherical region bounded by $L$ no longer lies within a trapped region and the dust sphere bounces (this meets general expectations described in \cite{BenAchour:2020gon}). Fourth, a shock wave forms after the bounce, the shock-wave front briefly moves beyond the inner horizon and then the inner horizon rapidly moves outwards and catches up, at which point the shock and the inner horizon move outwards together. Finally, the shock wave eventually reaches the outer apparent horizon located at $x = R_S$, at this time the inner and outer horizons meet and annihilate, and the black hole disappears.

An important point is that due to the gravitational shock wave that forms after the bounce, there is a discontinuity in the gravitational field, namely in the space-time metric. This means that the $x=L$ surface has different properties with respect to the inner and outer metrics. In particular, during most of the fourth stage mentioned above, the $x=L$ surface is timelike according to the inner space-time metric, but spacelike according to the outer metric. Further, when constructing the conformal diagram, different coordinate transformations will be required for the interior and exterior regions for stages four and five after the bounce (and these are discontinuous across the boundary); as a result the location of the boundary surface $x=L$ in the conformal diagram will not be the same with respect to the interior and exterior metrics. Due to this, we identify the two locations of the boundary surface and excise the region in the conformal diagram that lies between the location of the boundary with respect to the inner and outer metrics.

With the overview complete, we now revisit each of the stages described above in more detail, first determining the location of the apparent horizons, and then describing the trajectory of $L(t)$. We do not construct the conformal diagram rigourously through coordinate transformations of the metric, but rather provide a sketch by determining which horizons and trajectories are null, spacelike, or timelike.

\begin{figure}
\begin{center}
\includegraphics[width=0.5 \textwidth]{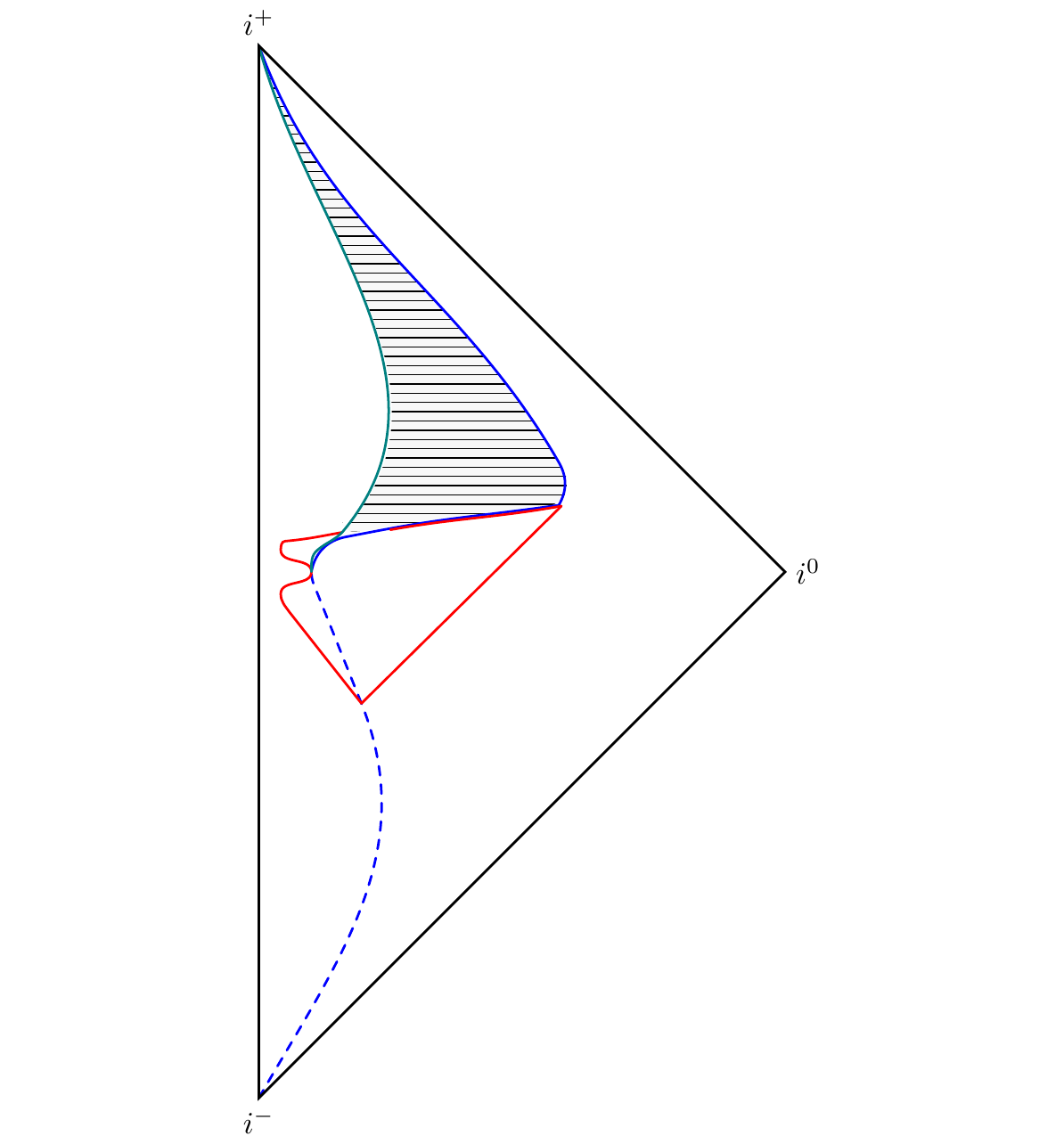}
\caption{Conformal diagram for the Oppenheimer-Snyder model showing the radius of the collapsing dust sphere (dashed line), the inner and outer apparent horizons (red lines) and the outgoing shock wave as seen from the exterior (dark blue line) and as seen from the interior (light blue line). The outgoing trajectories of the shock wave are identified, and the region in between (horizontal hatching) is excised from the conformal diagram.}
\label{figure:OS-Diag}
\end{center}
\end{figure}

In spherical symmetry, the location of the apparent horizons is given by the zeros of $\Theta_+ = |\nabla x|^2$ \cite{Faraoni:2016xgy}, which for the Painlev\'e-Gullstrand form of the metric (with $E^b=x$) gives
\beq
\Theta_+ = |\nabla x|^2 = 1 - (N^x)^2,
\ee
and recall that in the effective theory $N^x = -(x/\gamma \sqrt\Delta) \cdot \sin\beta \cos\beta$.

During collapse, the shift vector in the interior ($x \le L$) is given by the upper relation in \eqref{eq:confshift}, so for the interior region the zeros $x_{h,i}$ of $\Theta_+$ are located at
\beq \label{innerh}
x_{h,i}(t) = \f{3|t|}{2} + \f{2 \gamma^2 \Delta}{3|t|}.
\ee
Of course, there is only an apparent horizon in the interior if the location of this $x_{h,i}$ lies within $L = (\tfrac{9}{4} R_S t^2 + \gamma^2 \Delta R_S)^{1/3}$. Neglecting quantum gravity corrections, it is easy to verify that an apparent horizon appears at $t = -2 R_S / 3$ at the location $x_{h,i} = R_S$, and then moves inwards as $x_{h,i} = -3t / 2$, this is a faster rate than $L$.

Once quantum gravity corrections become important close to the bounce time $t=0$, the apparent horizon in the interior will cross over $x_{h,i}=L$ and enter the exterior region. The time this occurs can be approximated by assuming the quantum gravity corrections dominate, so the location of the apparent horizon is $x_{h,i} \approx 2 \gamma^2 \Delta / (6t)$ while $L \approx L_{\rm bounce} = (\gamma^2 \Delta R_S)^{1/3}$, implying a crossover time of $t_c \approx -(8 \gamma^4 \Delta^2 / 27 R_S)^{1/3}$. Substituting this back into $L(t)$ gives $L(t_c) \approx L_{\rm bounce} + (\gamma^4 \Delta^2 / 27 R_S)^{1/3}$.

Also note that the minimum radius of the apparent horizon in the interior is $x_{h,i} = 2 \gamma \sqrt\Delta$, which is reached at $t = - 2 \gamma \sqrt\Delta \, / 3$.

To recap, there is an apparent horizon inside the dust sphere during the collapse, between the times when $L$ reaches the radii $L \approx R_S$ and $L \approx L_{\rm bounce} + (\gamma^4 \Delta^2 / 27 R_S)^{1/3}$. Importantly, note that $\Theta_+ > 0$ for $x < x_{h,i}(t)$, so it is the region $x > x_{h,i}$ that is trapped, while the region $x < x_{h,i}(t)$ is not trapped; this shows that this apparent horizon is an inner horizon.

For the exterior, the shift vector during collapse is given by the lower relation in \eqref{eq:confshift}, so the zeros of $\Theta_+$ for the vacuum exterior region are located at the solutions of the following implicit equation for $x_{h,e}$,
\beq \label{vac-h}
x_{h,e} = R_S \left( 1 - \f{\gamma^2 \Delta R_S}{x_{h,e}^3} \right).
\ee
Once again, these apparent horizons are only present if $x_{h,e} > L(t)$. It is immediately clear that there will be an apparent horizon at $x_{\rm outer} = R_S$, neglecting quantum gravity corrections, once $L(t)$ passes the Schwarzschild radius. As expected, this is an outer horizon since $\Theta_+ < 0$ for $x < R_S$.

There is another solution to \eqref{vac-h} that gives a second apparent horizon at $x_{\rm inner} \approx L_{\rm bounce} + (\gamma^4 \Delta^2 / 27 R_S)^{1/3}$, whose location matches the radius where the interior horizon $x_{h,i}$ exits the surface $L$ of the dust sphere as described above. This is an inner horizon where $\Theta_+ > 0$ for $x < x_{\rm inner}$, again as expected.

In summary, during the collapse phase a pair of apparent horizons forms when $L$ reaches the radius $x \approx R_S$ (up to small Planckian corrections), with the outer horizon remaining at $R_S$ in the vacuum exterior, while the radius of the inner horizon decreases faster than $L$ (staying within the dust sphere) until it reaches $x_{\rm inner}$, at which time the inner horizon crosses outside the trajectory of $L$, enters the vacuum exterior region and stays at $x_{\rm inner}$.

After the bounce, a shock wave forms and moves outwards, as described in Sec.~\ref{s.os-collapse}. The outer horizon at $R_S$ will stay there until the outgoing shock $L$ reaches it a time $\sim M^2 / m_{\rm Pl}$ later, at which point it will disappear. On the other hand, the inner horizon will follow \eqref{innerh} and contract back to its minimum radius of $x_{h,i} = 2 \gamma \sqrt\Delta$ at $t = 2\gamma \sqrt\Delta / 3$ before expanding once more. For the short period of time that $L > x_{h,i}$, the dust located inside this region is moving outwards in a trapped region---this does not indicate a violation of the dominant energy condition, but rather is due to quantum gravity effects being large and significantly modifying the dynamics from what could be expected from classical general relativity. After $t = 2 \gamma \sqrt\Delta / 3$, the inner apparent horizon will rapidly move outward until it reaches the shock, at which point it will move in lockstep with the shock until they reach the outer apparent horizon (a time $\sim M^2 / m_{\rm Pl}$ later) and then the two apparent horizons will disappear. After this, even though the black hole is now gone the shock will continue to move outwards forever.

We note also that the dust is not superluminal as it emerges from the horizon. This may be seen by recalling that for the Oppenheimer-Snyder collapse model considered here, the dust ball is isomorphic to (a portion of) the FLRW space-time. As in the FLRW space-time, the dust is not superluminal---it is just co-moving with the region of space-time that is rapidly expanding after the bounce.

The trajectory of the two (outer and inner) apparent horizons, as described, is shown in the conformal diagram in Fig.~\ref{figure:OS-Diag}, the two apparent horizons are shown as red lines.

The other trajectory of interest in the conformal diagram is the path followed by $L(t)$, denoting the surface of the dust sphere during collapse, and the location of the shock wave after the bounce. The normal vector to the surface $x=L(t)$ is
\beq
n_\mu = -\left( \f{\dd L}{\dd t} \right) (\dd t)_\mu + (\dd x)_\mu,
\ee
and the sign of
\beq
g^{\mu\nu} n_\mu n_\nu = 1 - \left( \f{\dd L}{\dd t} \right)^2 - 2 N^x \left( \f{\dd L}{\dd t} \right) - (N^x)^2
\ee
will determine whether the trajectory of $L$ is timelike or spacelike.

During the collapse, using the solution \eqref{L-cont} for $L(t)$ and the metric \eqref{modmetric} with the shift vector \eqref{eq:confshift} evaluated at $x=L$, a direct calculation gives $n^\mu n_\mu = 1$ at all times---independently of whether the interior or exterior metric is used---showing that $L(t)$ follows a timelike trajectory during the collapse.

After the bounce, the norm of $n_\mu$ will depend on the metric that is used; due to the discontinuity in the gravitational field, using the interior metric or the exterior metric will give a different answer.

With respect to the interior metric,
\be \label{eq:intnorm}
(n^\mu n_\mu)_i = 1 - \left(\f{\dd L}{\dd t}\right)^2
+ \f{12 \, t \, L(t)}{9 t^2 + 4 \gamma^2 \Delta} \cdot \left( \f{\dd L}{\dd t} \right)
- \f{36 \, t^2 \, L(t)^2}{(9 t^2 + 4 \gamma^2 \Delta)^2},
\ee
while with respect to the exterior metric,
\be \label{eq:extnorm}
(n^\mu n_\mu)_e
= 1 - \left(\f{\dd L}{\dd t}\right)^2
-2 \, \sqrt{ \f{R_S}{L(t)} - \f{\gamma^2\Delta R_S^2}{L(t)^4} \, } \cdot \left( \f{\dd L}{\dd t} \right)
- \f{R_S}{L(t)} + \f{\gamma^2\Delta R_S^2}{L(t)^4}.
\ee
Note that in both cases $L(t)$ is given by the same solution to the differential equation \eqref{os-shock}. Clearly, if $n^\mu n_\mu > 0$, then the trajectory of $L(t)$ is timelike, while if $n^\mu n_\mu < 0$, then $L(t)$ is moving in a spacelike manner.

By numerically solving \eqref{os-shock} for $L(t)$, it can be verified that, according to the interior metric, $L(t)$ follows a trajectory that is almost always timelike, except for the brief interval after the bounce when $L > x_{h,i}$. On the other hand, the post-bounce trajectory of $L$ with respect to the outer metric is spacelike until $L$ reaches the outer apparent horizon at $x=R_S$, at which instant it is null and then immediately after becomes timelike.

This is shown in the conformal diagram in Fig.~\ref{figure:OS-Diag}, where the trajectory of $L(t)$ during collapse is given by the dashed blue line, while after the bounce the trajectory of $L(t)$ with respect to the interior and exterior metrics is shown in pale blue and dark blue respectively. As discussed above, there are two trajectories for $L(t)$ in the conformal diagram after the bounce, this is due to the discontinuity in the metric across the shock wave located at $L(t)$ after the bounce: different coordinate transformations are required for the interior and exterior regions to construct the conformal diagram, so the common boundary of the interior and exterior does not have the same location in the conformal diagram. The conformal diagram is made whole by identifying the two locations of the boundary, and excising the portion of the conformal diagram in between (marked with a horizontal hatching).

To understand this identification in the conformal diagram, consider an infalling particle that reaches the dark blue solid line in Fig.~\ref{figure:OS-Diag} corresponding to the location of the shock as seen from the exterior. When the particle reaches a point on the dark blue line, it also lies on a point on the light blue line since these lines are physically the same and therefore identified (recall that the hatched region is excised from the conformal diagram). The particle would then continue to move inwards, above the light blue line in the region inside the shock wave.

In summary, the qualitative picture combining the locations of the apparent horizons and also $L(t)$ is depicted in the conformal diagram shown in Fig.~\ref{figure:OS-Diag}: the star surface collapses to form a black hole with the formation of an outer null horizon at $x = R_S$ (neglecting tiny Planckian corrections to the location of the outer apparent horizon), together with an inner dynamical horizon; a bounce occurs when $L(t) = L_{\rm bounce} = (\gamma^2 \Delta R_S)^{1/3}$ and an outgoing gravitational shock wave forms; the shock wave slowly moves outward, and when the inner and outer horizons meet and annihilate the black hole ceases to exist.

We note that the conformal diagram constructed here is specifically for Oppenheimer-Snyder collapse. In Sec.~\ref{s.numerics}, we present the conformal diagram inferred from numerical solutions for a different family of initial data corresponding to the collapse of dust with a Gaussian radial density profile, see Fig.~\ref{figure:cdiag}. Although the conformal diagrams have some differences, the important qualitative features are similar: there is a non-singular bounce, a gravitational shock wave, and the eventual disappearance of the black hole when the inner and outer apparent horizons meet and annihilate after a time $\sim M^2 / m_{\rm Pl}$.

\section{Numerical solutions}
\label{s.numerics}

As discussed in the previous section, shell crossings are a feature of the effective dynamics, and evolution beyond the point where characteristic curves cross requires finding weak solutions; a numerical approach is generally needed for this. In this section we review and apply the well-known Godunov method to this problem, this is one of a family of numerical techniques available for solving non-linear flow equations.

\subsection{Godunov method}

The Godunov method is a numerical approach for solving differential equations that are non-linear conservation laws of the form $\partial_t\rho(x,t) + \partial_x j(\rho)=0$; see, e.g., \cite{Leveque} for an in-depth discussion of this and related algorithms.

To use this method we first write the evolution equation~\eqref{eomb} by defining 
\be \label{def-Bm}
B(x,t) = x \, b(x,t) \quad \text{and} \quad m(x,B) = \f{x^3}{2\gamma^2 \Delta} \sin^2 \left( \f{B}{x^2} \right).
\ee
This gives the desired form of a conservation law
\be \label{Bigb}
\partial_t B(x,t) + \partial_x m(x,B) = 0\ ;
\ee
the current in this equation is the mass function $m(x,t)$, which is related to the dust density \eqref{q-density}:
\be
m(x,t) = 4\pi \int_0^x \dd \tilde x ~ r^2 \rho(\tilde x,t).
\ee
It is also evident from \eqref{def-Bm} that the current has an explicit $x$ dependence. 

Integrating \eqref{Bigb} over a spatial interval $x_L \leq x \leq x_{R}$ and time interval $t_{1} \leq t \leq t_{2}$ gives the integral form of the conservation equation \eqref{Bigb},
\be \label{int_cons}
\int_{x_{L}}^{x_{R}} \dd x\,\, B(t_{2},x) = \int_{x_{L}}^{x_{R}} \dd x \,\, B(t_{1},x) - \int_{t_{1}}^{t_{2}} \dd t \,\, \Big[ m(x,B) \Big]^{x_{R}}_{x_{L}}.
\ee

In either form the conservation equation is non-linear and cannot be solved analytically (except for a few especially simple configurations like thin shells and the Oppenheimer-Snyder model). Finite-volume methods like the Godunov scheme are based on the approximation that at each time step, the field is piecewise constant in every spatial cell $j$ of width $\delta x$; that is, the field at the point $(x,t_n)$ is defined by the spatial average over the cell
\be
B^{n}_j = \frac{1}{\delta x} \int^{x_j + \tfrac{1}{2} \delta x}_{x_j - \tfrac{1}{2} \delta x} \dd x \,\, B(x, t_{n}),
\ee
and this average value is then assigned to each point in the interval,
\be
B(x, t_n) = B_j^n, \qquad x \in (x_j - \tfrac{1}{2} \delta x, x_j + \tfrac{1}{2} \delta x).
\ee
In this way, the field $B(x,t)$ is taken to be piecewise constant.

This discretization makes a numerical integration of the equation \eqref{int_cons} possible on a space-time lattice of spacings $\delta t$ and $\delta x$,
\be \label{update}
B^{n+1}_j = B^{n}_j - \frac{1}{\delta x} \int_{t_{n}}^{t_{n+1}} \dd t \,\, \Big[ m(x,B) \Big]^{x_j + \tfrac{1}{2} \delta x}_{x_j - \tfrac{1}{2} \delta x}.
\ee
This form of the equation shows that $B^{n+1}_j$ can be calculated by adding the flux $m(x,B)$ across the boundaries $x_j + \tfrac{1}{2} \delta x$ and $x_j - \tfrac{1}{2} \delta x$ to the initial value $B^n_j$. However the fact that the time integral is from $t_n$ to $t_{n+1}$ means that the method is implicit and the net flux in the integrand requires careful construction. This is accomplished by the Godunov method. 

The key insight underlying the Godunov method is that, given the initial conditions of a piecewise constant field $B$, it is possible to solve exactly for the flux functions $m(x_j + \tfrac{1}{2}\delta x, B)$ and $m(x_j - \tfrac{1}{2}\delta x, B)$, at least for a short period of time.

The basic idea is illustrated by focusing on the flux through the boundary $x_{bd} = x_j + \tfrac{1}{2} \delta x$: if $B$ is moving outwards, the flux across the boundary is given by $m(x_{bd}, B_j^n)$; if the field is moving inwards the flux is given by $m(x_{bd}, B_{j+1}^n)$. These simple forms are due to the piecewise constant prescription for $B$ within the cell $[x_j - \tfrac{1}{2} \delta x,x_j + \tfrac{1}{2} \delta x]$. A similar consideration applies to the cell boundary at $x_j - \tfrac{1}{2} \delta x$.

It remains to determine the direction in which the flux moves. This requires the velocity of $B$, given by the derivative of the current function $m$ in \eqref{Bigb} with respect to $B$,
\beq \label{def-v}
v = \partial_B m = \f{x}{\gamma^2 \Delta} \sin \left(\f{B}{x^2}\right) \cos \left(\f{B}{x^2}\right);
\ee
$v$ depends on both the position $x$ and the value of the field $B$.

There are four possible types of $B$-field velocity configurations at each boundary $x_{bd}$ between two lattice cells. Denoting quantities on the left and right of the boundary by the subscripts $L$ and $R$ respectively, the possibilities are:
\begin{enumerate}
\item[(i)] $v_L\ge 0\ \text{and}\ v_R \ge 0$: field moving right/outwards;
\item[(ii)] $v_L\le 0\ \text{and}\ v_R\le 0$: field moving left/inwards;
\item[(iii)] $v_L \ge 0 \ge v_R$: shock wave;
\item[(iv)] $v_L < 0 < v_R$: rarefaction wave.
\end{enumerate}
The first two cases are simple as they have velocities in the same direction. The third case corresponds to a shock wave with speed $v_s = [m]/[B]$ \eqref{rh-cond}, with its sign determining whether the shock moves outwards (positive $v_s$) or inwards (negative $v_s$), and flux $m(x_{bd}, B_j^n)$ or $m(x_{bd}, B_{j+1}^n)$ respectively. The fourth case is a rarefaction wave, the corresponding flux is obtained by taking $m(x_{bd}, B_s)$ where $B_s$ is the stationary value of $B$ where $v=0$. From \eqref{Bigb} and \eqref{def-v}, $v=0$ for $B = \pi x^2 / 2$, for which $m(x_{bd}, B_s) = x_{bd}^3 / (2\gamma^2 \Delta)$.

By looking at the explicit forms in terms of $B$ of the current $m(x,B)$ in \eqref{def-Bm} and the velocity $v(x,B)$ in \eqref{def-v}, the flux for these four cases are summarized by
\beq \label{godunov}
F(B_j^n, B_{j+1}^n, x_{bd}) =
\begin{cases} 
  \text{min}_{B_j^n \leq B \leq B_{j+1}^n} \Big( m(x_{bd}, B) \Big) & {\rm for}~ B_j^n \le B_{j+1}^n, \\
  \text{max}_{B_{j+1}^n \leq B \leq B_j^n} \Big( m(x_{bd}, B) \Big) & {\rm for}~ B_j^n > B_{j+1}^n.
\end{cases}
\ee
The min and max refer the minimum/maximum value attained by $m(x_{bd},B)$ for $B$ in the given interval. This is the general Godunov scheme for the flux; it holds also for a non-convex flux function ($m$ in this case); i.e., for either sign of $\partial_B^2 m$. Here, due to the relatively simple (although non-convex) form of $m$, for the interval $B \in [-\pi x_{bd}^2, 0]$ the minimum is always one of the two endpoints, while the maximum is either one of the two endpoints or the stationary point $m(x_{bd}, B_s) = x_{bd}^3 / (2\gamma^2 \Delta)$ if the stationary point $B_s$ lies between $B_L$ and $B_R$.

With this prescription of the flux function, discrete evolution equation takes the explicit form
\be \label{discreteevol}
B_j^{n+1} = B_j^n - \frac{\delta t}{\delta x} \Big[ F( B_{j+1}^n, B_j^n, x_j + \tfrac{1}{2} \delta x ) - F( B_j^n, B_{j-1}^n, x_j - \tfrac{1}{2} \delta x ) \Big],
\ee

For boundary conditions, since $B = xb$ we impose that $B(x=0) = 0$, and we also assume that there is no infalling matter coming from beyond the outermost lattice point $x_{\rm last}$ by assuming that $\dot B(x_{\rm last}) = 0$, recall that $\dot B = 0$ in vacuum as seen in \eqref{eom-beta-rho}.

Lastly, the discrete evolution scheme is stable provided $\delta t$ is chosen small enough to satisfy the Courant-Friedrich-Lewy condition that $\delta t < \delta x / |v_{\rm max}|$ where $|v_{\rm max}|$ is the maximal speed $v$ at any boundary $x_{bd}$ for the given time step. In the numerical code, we determine $\delta t$ dynamically by finding the maximum characteristic speed $v_{\rm max}$ at each time step and using this to fix $\delta t$ to an appropriate value.
 
The \textsc{matlab} code we used is available online \cite{code}.

\subsection{Results}

We used the algorithm described above to generate numerical solutions for Gaussian and hyperbolic tangent initial density profiles given by 
\bea
\rho_0^G(x) &=& \exp \Big( -(x-x_0)^2/\sigma^2 \Big), \\
\rho_0^T(x) &=& 1 + \tanh \Big( -(x-x_0)/\sigma \Big),
\eea
that we use to construct an initial mass function
\be \label{init-m}
m(x,t=0) = M \, \frac{\int_0^x \dd \tilde x \ \rho_0(\tilde x)}{\int_0^\infty \dd \tilde x\ \rho_0(\tilde x) },
\ee
where $M$ denotes the total mass. We considered $M$ ranging from $5 m_{\rm Pl}$ to $\sim 200 m_{\rm Pl}$. Inverting \eqref{def-Bm} gives the initial profile for $B$,
\beq
B(x,t=0) = - x^2 \arcsin \left( \sqrt{ \f{2 \gamma^2 \Delta \, m(x, t=0)}{x^3} } \right),
\ee
and we evolve $B$ using the Godunov method described above. The negative sign of the square root corresponds to a density profile that is initially contracting, this is the case of interest for gravitational collapse. 

\begin{figure}[t]
\begin{center}
\includegraphics[width=\textwidth]{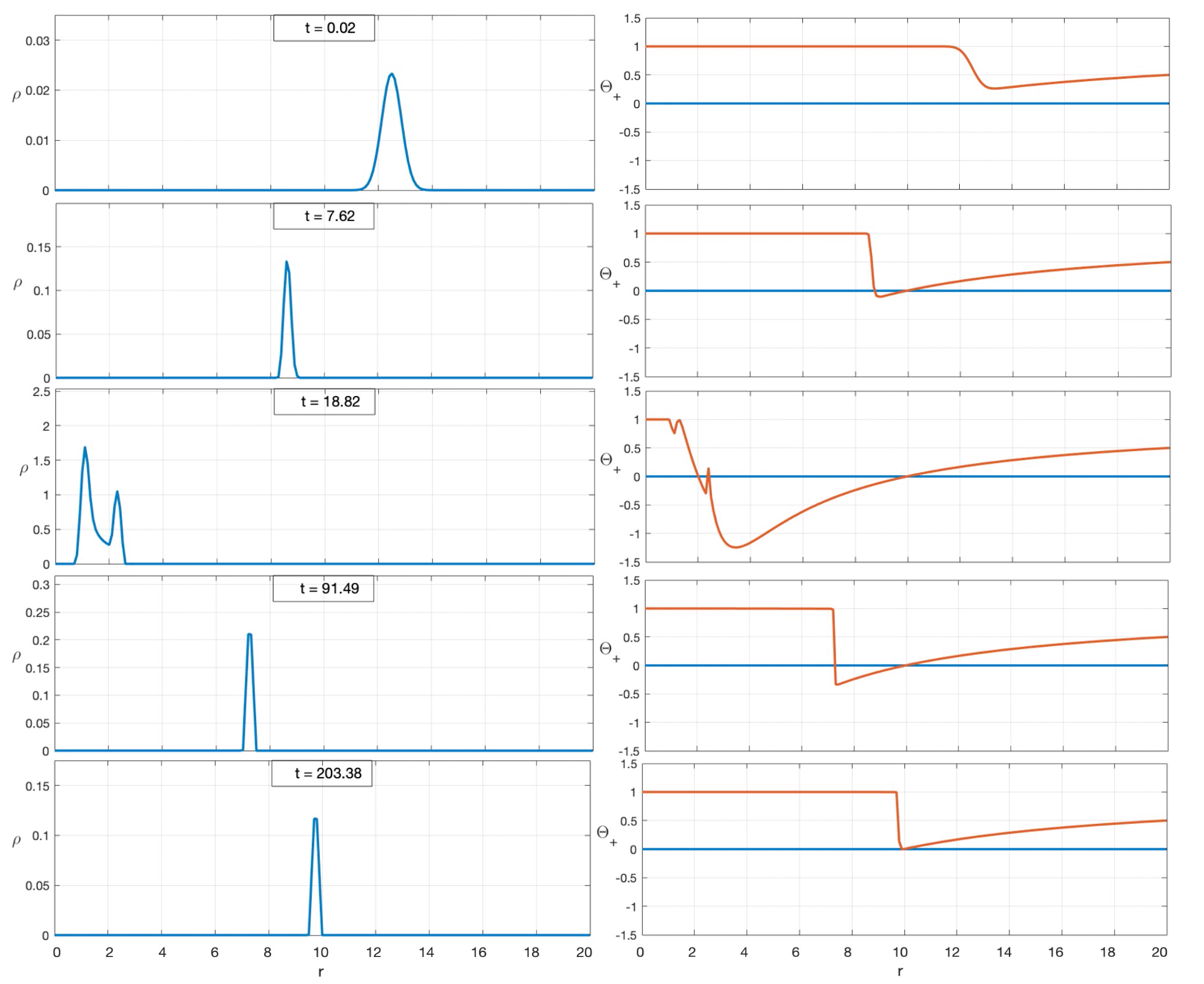}
\end{center}
\caption{Frames from a numerical simulation of black hole formation and evolution for the Gaussian initial density profile with $M = 5 m_{\rm Pl}$; the left column is the energy density $\rho$; the right column is the function $\Theta_+$; its zeros give the locations of apparent horizons. The middle frame shows part of the bounce. Axes are in Planck units and $\gamma^2 \Delta=1$.}
\label{frames-g}
\end{figure}

\begin{figure}[t]
\begin{center}
\includegraphics[width=\textwidth]{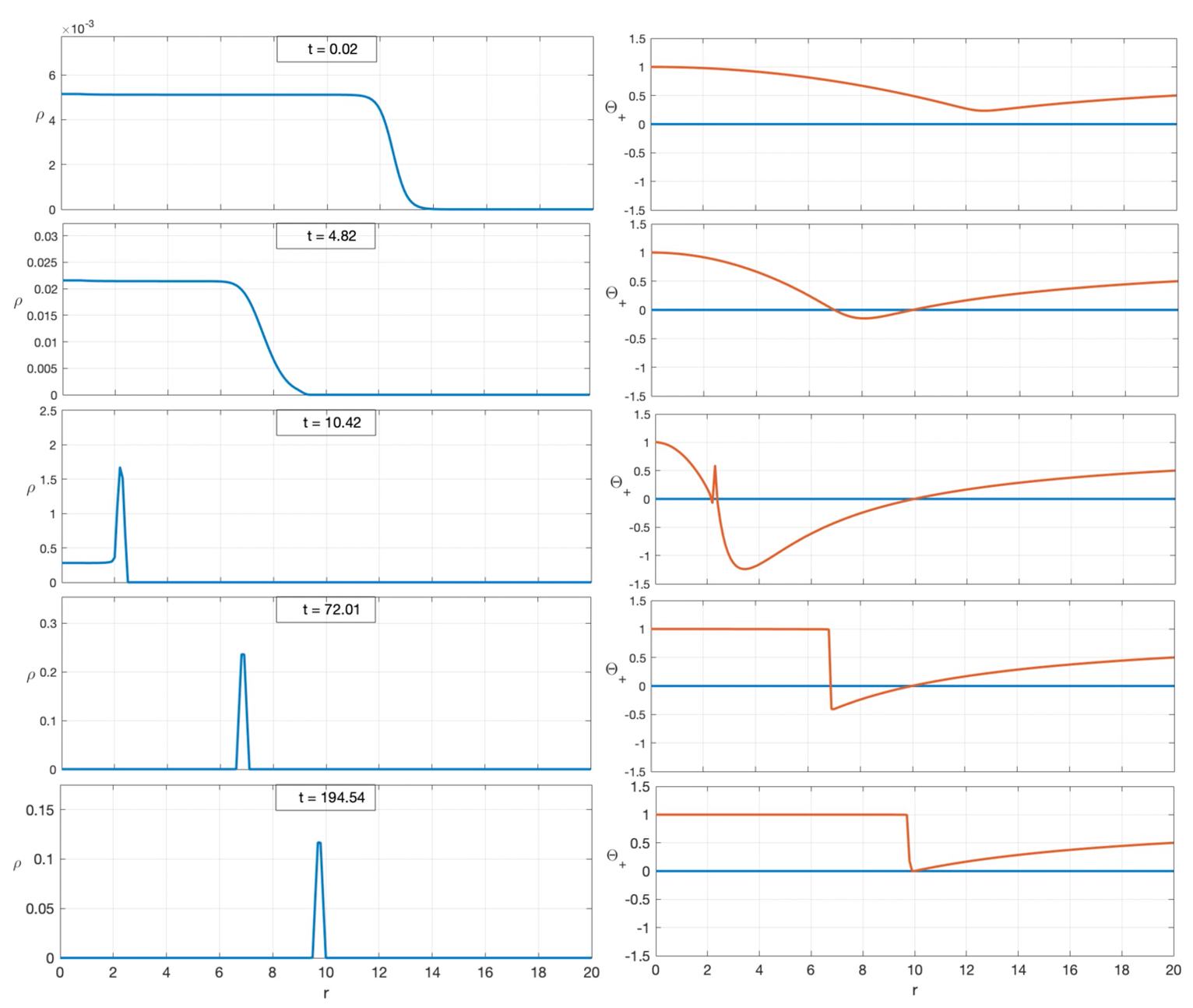}
\end{center}
\caption{Frames from a numerical simulation of black hole formation and evolution for the Gaussian initial density profile with $M = 5 m_{\rm Pl}$; the left column is the energy density $\rho$; the right column is the function $\Theta_+$; its zeros give the locations of apparent horizons. The middle frame shows part of the bounce. Axes are in Planck units and $\gamma^2 \Delta=1$.}
\label{frames-tanh}
\end{figure}

\begin{figure}[t]
\begin{center}
\includegraphics[width=\textwidth]{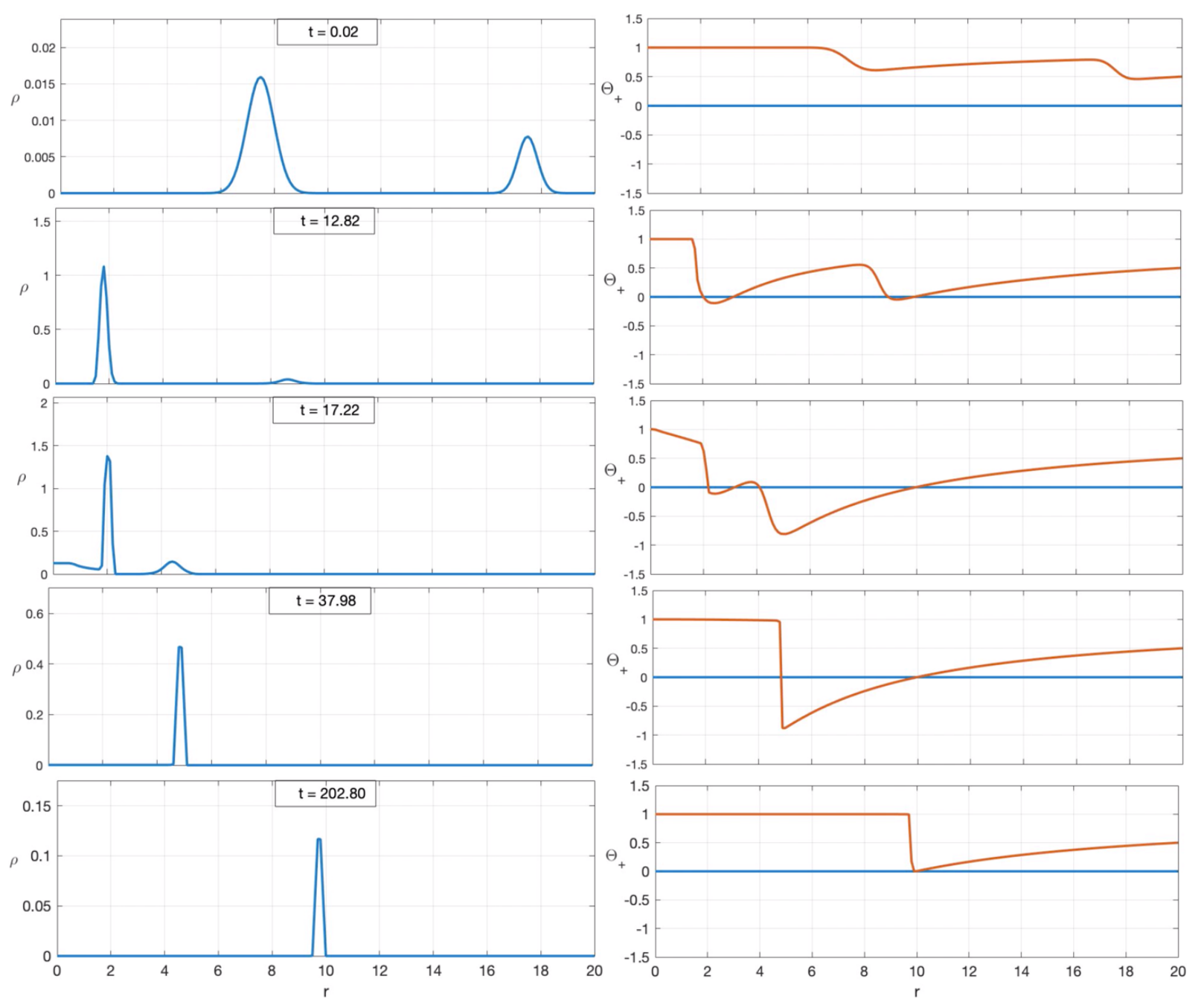}
\end{center}
\caption{Frames from a numerical simulation of black hole formation and evolution for the initial density profile $\rho_0(x) = \exp[ -2 (x - 7.5)^2 ] /5 + \exp[ -4 (x - 17.5)^2] /10$ in \eqref{init-m} with $M = 5~m_{\rm Pl}$; the left column is the energy density $\rho$; the right column is the function $\Theta_+$; its zeros give the locations of apparent horizons. The middle frame shows part of the bounce. Axes are in Planck units and $\gamma^2 \Delta=1$. These frames are also shown in the companion letter \cite{Husain:2021ojz}.}
\label{frames-2}
\end{figure}

Time frames from a typical simulation of Gaussian data for $M=5~m_{\rm Pl}$ are shown in Fig.~\ref{frames-g}. The left column displays the evolving density profile at the displayed times, and the right column is the function
\beq
\Theta_+ (x,t) = |\nabla x|^2 = 1-(N^x)^2 = 1- \frac{x^2}{4}\sin^2\left(\frac{2B}{x^2} \right),
\ee
the roots of $\Theta_+$ give the locations of the evolving apparent horizons \cite{Faraoni:2016xgy}. A number of features are apparent in Fig.~\ref{frames-g}: as the density profile moves inward, it compresses to become a sharp pulse; the outer horizon forms at the Schwarzschild radius $R_S = 2GM = 10 \lp$ as the pulse crosses this radial location; the inner horizon moves inward until the matter bounces, and then moves outward; finally in the last frame the inner and outer horizons merge and disappear as the outgoing shock wave exits the horizon. These general features can be seen for all choices of $M$ (except very small $M < 8 \gamma \sqrt\Delta / \sqrt{27} G$ for which a horizon never forms \cite{Kelly:2020uwj}, although the bounce and shock formation still occur).

The collapse for hyperbolic tangent initial data is somewhat different, but there is also a bounce and an outgoing shock, this is shown in Fig.~\ref{frames-tanh}. The outgoing evolution is quite similar to that for Gaussian initial data: the density becomes sharply peaked at the shock, and the inner horizon moves outwards until it meets the outer horizon. These general features occur for all hyperbolic tangent initial data with $M > 8 \gamma \sqrt\Delta / \sqrt{27} G$.

The evolution of two Gaussian profiles is shown in Fig.~\ref{frames-2}. Again the main features are qualitatively similar: horizon formation, non-singular bounce, shock formation, and eventual disappearance of the horizons. There are two notable points for the double pulse results: the second ingoing profile does not cause the outgoing shock wave to recollapse, and the mass function remains bounded at the inner horizon. These features show a robustness of the results to perturbations. 

\begin{figure}[t]
\begin{center}
\includegraphics[width=0.45\textwidth]{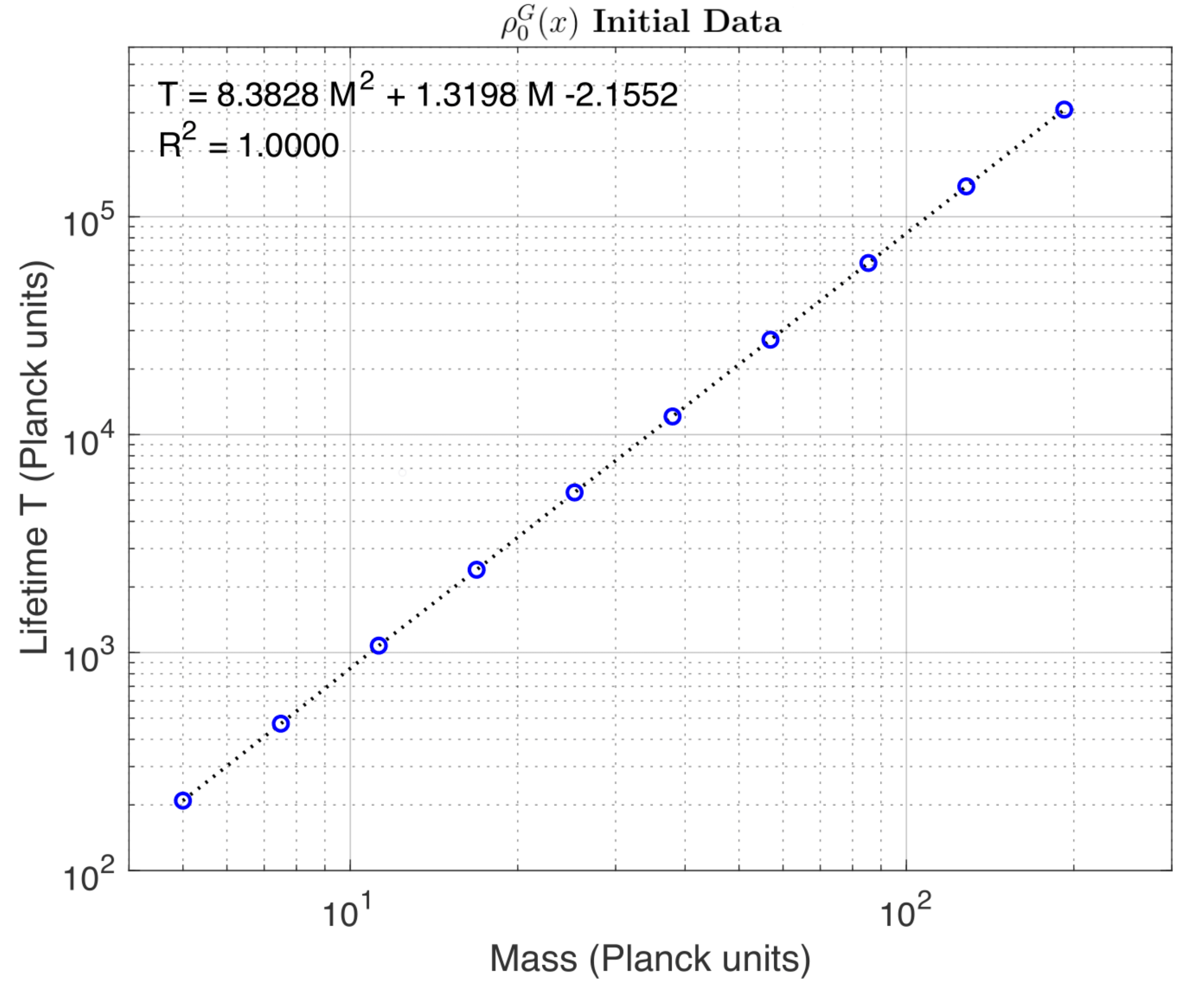}
\includegraphics[width=0.45\textwidth]{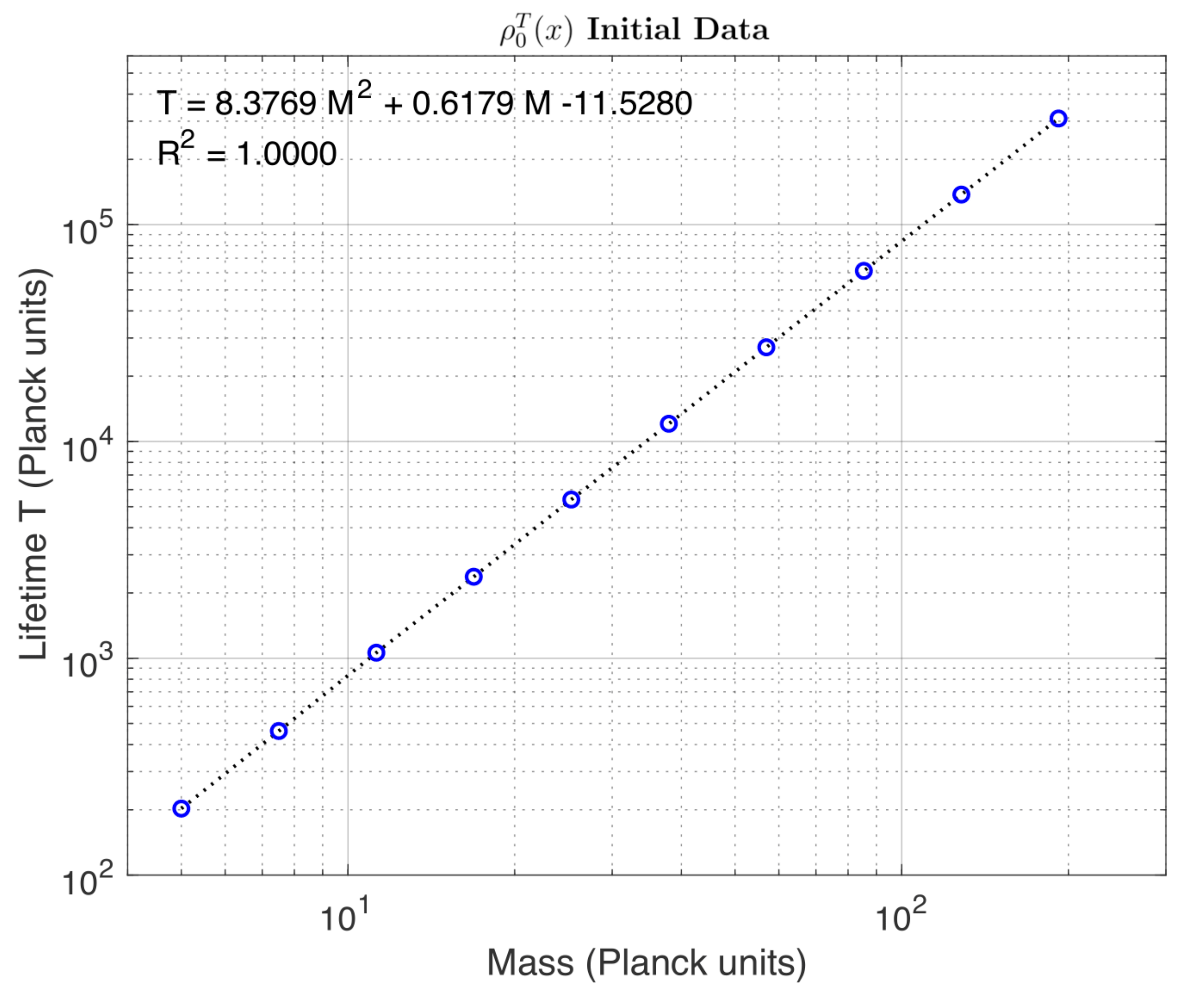}
\end{center}
\caption{Black hole lifetime $T$ as a function of data mass $M$ for Gaussian (left) and hyperbolic tangent (right) initial density profiles. The leading order dependence is $T = 8 \pi M^2 / 3 + {\cal O}(M)$ to an excellent approximation. Axes are in Planck units with $\gamma^2\Delta = 1$. These plots are also shown in the companion letter \cite{Husain:2021ojz}.}
\label{life-fig}
\end{figure}

We also calculated the black hole lifetime $T$ from numerical simulations. This is the time between the formation and disappearance of its outer apparent horizon. It is natural to calculate this using the proper time of a distant observer who observes a light ray emitted shortly before the collapse forming the black hole (for example, from a supernova of a collapsing star), and another light ray emitted by the shock wave once it has exited the outer apparent horizon. A short calculation shows that this proper time interval, as measured by a distant observer, equals the Painlev\'e-Gullstrand coordinate time between the formation and disappearance of the outer apparent horizon \cite{Kelly:2020lec}.

In our simulations, we numerically record the times at which the outer horizon appears and disappears, the results are shown in Fig.~\ref{life-fig}. The best fit is shown in each of the plots, in both cases the leading order behaviour is
\be
T \approx \frac{8\pi}{3} M^2 + {\cal O}(M), 
\ee
in units where $G=\hbar=c=1$, and having set $\gamma^2 \Delta = 1$ for the numerics.

This result is in good agreement with the expectation from analytic calculations in the Oppenheimer-Snyder and thin shell models \eqref{life-os} and \eqref{life-ts}. This not surprising: numerical simulations show that soon after the bounce, the dust density can be approximated by a thin shell, so the thin shell calculation gives the correct leading order contribution to $T$.

The predicted black hole lifetime $T \sim M^2 / m_{\rm Pl}$ is very long in astrophysical terms, but short compared to the lifetime of a black hole as predicted by standard Hawking evaporation calculations. We comment below in Sec.~\ref{s.impl} on the implications of this result for the information loss problem.

This completes our description of numerical results. We now provide several comments and describe the conformal diagram suggested by our simulations.

The outgoing shock wave may be viewed in some ways as ``white hole." Indeed, it has been argued that quantum gravity might generate a transition from a black hole to a white hole \cite{Haggard:2014rza}. But white holes are known to be unstable to infalling matter in general relativity; under small perturbations they recollapse and subsequently form a black hole \cite{Eardley:1974zz, Barcelo:2015uff}. Our results are similar in spirit in that a singularity is replaced by a bounce, but different in the important detail that a shock wave is not a white hole. Our simulation of the double Gaussian provides numerical evidence that the outgoing shock wave is stable to an ingoing perturbation. This distinguishes our result from the black to white hole transition ideas.

Another instability in classical general relativity is mass inflation \cite{Poisson:1989zz, Ori:1991zz, Husain:1994xa}. This is the observation that the mass function grows without bound at an inner Cauchy horizon under time-dependent perturbations. On this horizon, infalling radiation is infinitely blueshifted. Therefore, in generic collapse the expectation is that backreaction would produce a curvature singularity at an inner horizon. This effect has also been observed in some non-singular black hole models in loop quantum gravity \cite{Brown:2011tv}. 
 
There is no mass inflation evident in our simulations. While the curvature is not well-defined at the shock due to the jump discontinuity, the density remains finite at all times and the total mass in the space-time is dynamically conserved.

To summarize these points, our numerical simulations show that there is no recollapse of the shock wave, and there is no mass inflation. 

Our last observation is the determination of a conformal diagram. The diagram corresponding to our numerical solutions is qualitatively similar to that for the effective Oppenheimer-Snyder model described in Sec.~\ref{s.conf}. The main difference is during the collapse phase, where some features of the conformal diagram depend on the initial density profile. For the hyperbolic tangent data, the conformal diagram is essentially identical to the one derived analytically for the Oppenheimer-Snyder collapse model shown in Fig.~\ref{figure:OS-Diag}. 

\begin{figure}
\begin{center}
\includegraphics[width=0.25\textwidth]{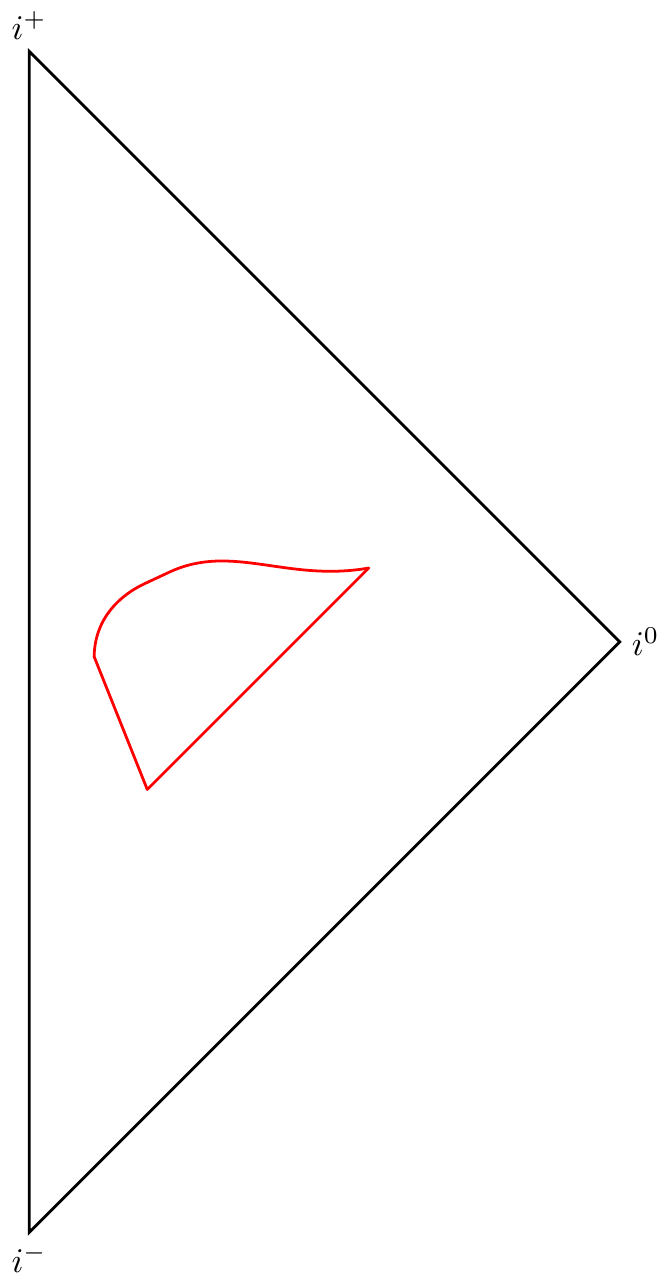} $~~~~$
\includegraphics[width=0.25\textwidth]{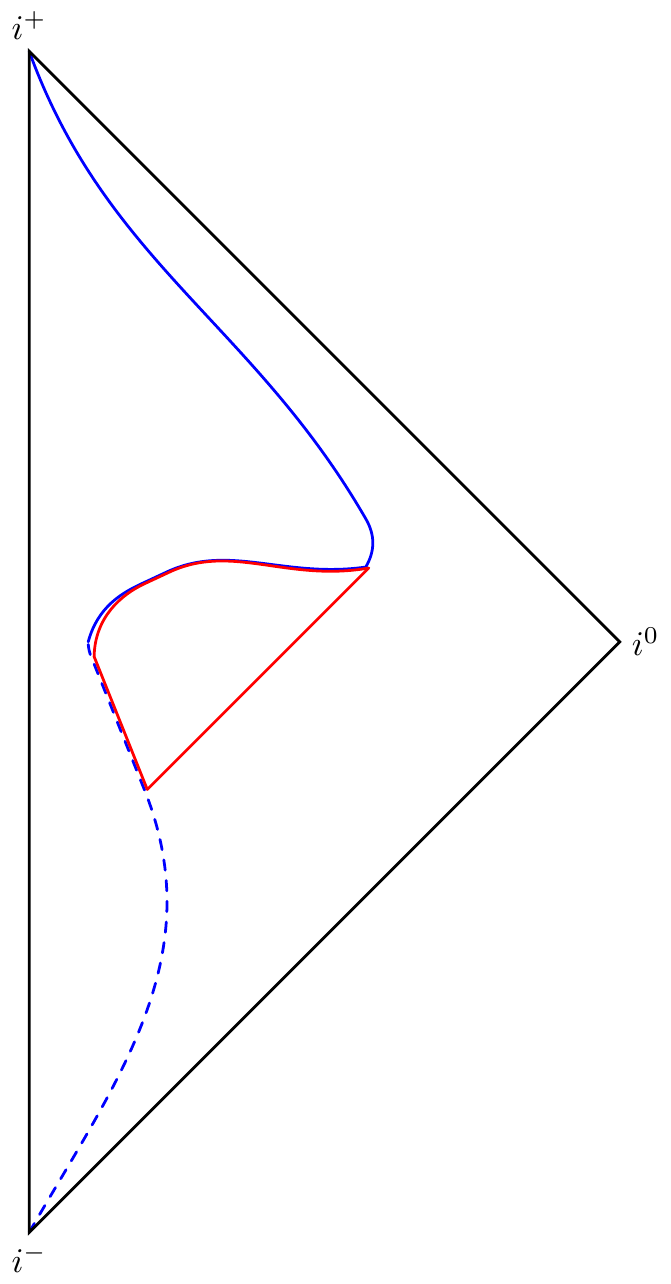} $~~~~$
\includegraphics[width=0.25\textwidth]{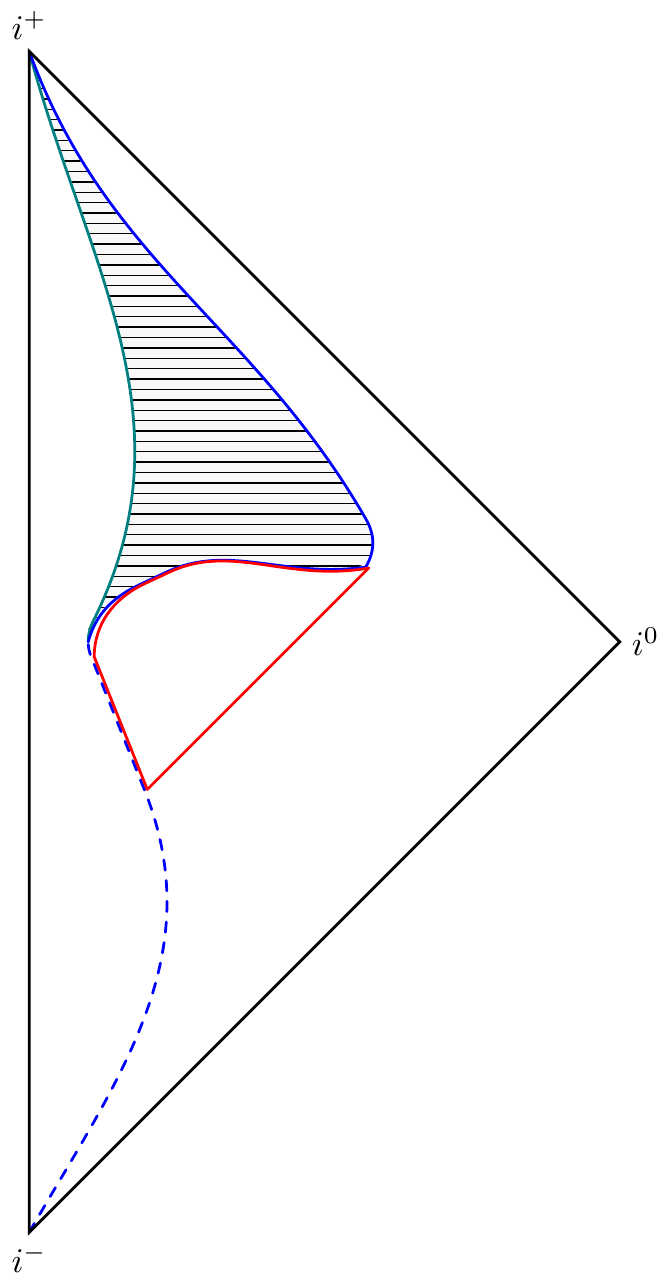}
\caption{The conformal diagram is shown in three frames. The left frame shows the apparent horizons in red; the outer horizon is null. The middle frame also includes the location of the collapsing Gaussian wave packet with the dashed blue line, and the solid blue gives the outgoing shock wave as seen by an outside observer, the shock wave exits at the point where the inner and outer horizons meet. The third frame illustrates the discontinuity in the metric at the shock wave---its trajectory is timelike from the perspective of the interior (shown in cyan), while it is spacelike with respect to the exterior metric until it exits the outer horizon at which time it becomes timelike; the hatched region is excised and the inner and outer shock wave trajectories are identified. The rightmost conformal diagram is also shown in the companion letter \cite{Husain:2021ojz}.}
\label{figure:cdiag}
\end{center}
\end{figure}

For Gaussian initial data, the conformal diagram is shown in Fig.~\ref{figure:cdiag}. It is slightly different in two ways. Firstly, the trajectory of the star surface is replaced by a packet of dust trajectories, and there may be multiple dynamical inner horizons depending on the type of initial data chosen (for example, if there are multiple Gaussian packets in the initial dust profile). Secondly, the inner horizon moves differently during the collapse phase: it moves inwards more slowly, following the Gaussian dust density. Despite these two points of difference, the main features remain the same: a collapse during which inner and outer horizons are formed, a non-singular bounce, formation and outward evolution of a shock wave, and the eventual disappearance of the inner and outer horizons when they meet.

As for the conformal diagram for the Oppenheimer-Snyder model, the trajectory of the shock wave can be calculated with respect to the inner space-time metric, or the outer one, with different results. This is why two trajectories are shown for the shock wave; these curves are identified in the conformal diagram and the hatched region in between is excised.

\section{Implications for black hole physics}
\label{s.impl}

We discuss here potential consequences of shock wave formation, and the implications of the predicted black hole lifetime for the information loss problem. 

\subsection{Shock wave formation and quantum geometry domain walls}

All our simulations show shock wave formation, this appears to be a generic feature of any initial data profile. For some initial profiles, a shock wave may form during the collapse phase; these are the cases that may be artificially avoided by imposing restrictions on initial data as in \cite{Szekeres:1995gy, Hellaby:1985zz, Booth:2005ng, Giesel:2009jp, Giesel:2021dug}. No matter the initial conditions, characteristic curves from the $\rho \neq 0$ region and the exterior vacuum region cross after the bounce.

Thus, in every case, including the double-peaked Gaussian profiles, an outgoing shock wave eventually emerges from the Schwarzschild radius signaling the end of the black hole phase. Although our simulations were limited to masses up to $\sim 200~m_{\rm Pl}$, the shock wave is expected to emerge from the outer horizon for data of any mass since there is no mechanism in the effective equation for the current to change sign during the outgoing phase. (Although the shock wave solutions we find may be viewed in some respects to be similar to a white hole solution, their properties are quite different: the shock wave solution is stable to infalling matter whereas a white hole solution is not \cite{Eardley:1974zz}.)

The occurrence of a shock wave has not been noticed in previous work, likely because many of the earlier studies focused on vacuum solutions or only included simple models for matter with a finite number of degrees of freedom (as opposed to a field theory with local degrees of freedom like the LTB space-time). Still, it is important to ask whether the shock wave is a robust prediction, or merely the consequence of some restriction, for example a gauge choice that may fail. As such, it would be useful to extend this work by relaxing the gauge conditions (namely the areal and dust-time gauges), but this generalization lies beyond the scope of this paper and is left for future work. Instead, we present some general arguments supporting the formation of a shock wave in a non-singular bouncing black hole in LQG.

Firstly, the bounce of the dust ball is not surprising---the region inside the collapsing dust sphere is locally similar to a contracting cosmological space-time (in fact, for the Oppenheimer-Snyder model this is an exact isomorphism), and these are known to bounce in loop quantum cosmology \cite{Ashtekar:2011ni}. Secondly, the vacuum region outside the dust sphere has no local degrees of freedom (at least in the spherically symmetric case we consider here) and therefore cannot evolve in the absence of matter; for example, in the marginally-trapped LTB space-times we consider here, by combining \eqref{eomb} and \eqref{q-density} it is easy to show that $\dot b=0$ when $\rho=0$. With these two ingredients, we expect a bounce inside the collapsing dust ball, but not outside. This causes a discontinuity to form, and it is the source of the shock wave.

Another way to see this is to consider an initial configuration with a dense core and a more dilute outer region; the core will bounce first and  collide with the outer region which is still collapsing; this causes a shell-crossing and thereby the formation of a shock wave. More generally, if there is an effective description of the space-time, it will presumably be governed by a wave equation that can reasonably be expected to be non-linear; solutions to non-linear wave equations are typically weak solutions, often including shock waves (especially solutions corresponding to highly energetic phenomena). Given this general expectation, it has been argued that it is necessary to allow weak solutions for LTB space-times even in classical general relativity \cite{Nolan:2003wp, Lasky:2006hq}. It is therefore not surprising that weak solutions are important also for effective metrics with quantum gravity corrections, as we find here. These arguments suggest that shock waves are a robust feature of quantum black holes formed from matter collapse, at least in LQG and possibly in other approaches to quantum gravity that resolve the black hole singularity.

There are different perspectives one can take when studying the shock waves found here. The simplest is to view the shock as a weak solution of the non-linear wave equation, as we have done. Another complementary perspective that may be valuable is to view the shock wave as a domain wall in space-time.

The late-time post-bounce effective solution has two regions, the inside region (lying within the shock wave), and the outside region. Since the shock wave is slowly moving outwards, the boundary between the two regions is also dynamical. Each region is separately well described by an effective line element, with a discontinuity across the shock. Taking seriously the perspective of LQG that a classical geometry should emerge from the coarse-grained description of many microscopic Planck-scale quanta of geometry, we propose the interpretation that the two regions are in different phases of the underlying quanta of geometry, with the shock wave being a domain wall separating the two (and carrying a non-zero energy density). From this perspective, the effective dynamics may have a thermodynamic interpretation, and the bounce can be understood as creating a domain wall separating two geometric phases.

At late times after the bounce, nearly all the dust accumulates on the shock wave, with the result that the interior region is nearly Minkowski and the exterior is nearly classical Schwarzschild. These are two vacuum solutions of classical general relativity, presumably corresponding to different microscopic configurations of the fundamental quanta of geometry. As the shock wave (or domain wall) moves outwards, the outside Schwarzschild vacuum geometry slowly relaxes to the Minkowski vacuum geometry. From a quantum geometry perspective, the shock wave is a domain wall that separates two different phases of vacuum quantum gravity.

\subsection{Observable consequences}

It is natural to expect that a shock wave emerging from what was a black hole could have significant observational consequences. The black hole lifetime result $T \sim M^2 / m_{\rm Pl}$ suggests that (dust) black holes of $\sim 10^{-8}$ solar masses have a lifetime of approximately the current age of the universe. If formed in the very early universe, such black holes would be on the verge of disappearing via shock wave emission at the present time. 

A more realistic collapse model will include several matter fields, not just dust. If the general picture obtained here continues to be applicable, then photons (for example) would be part of the shock wave, and be unable at first to move outward any faster due to the outer apparent horizon. After the shock wave exits the horizon, photons would no longer be trapped and could move outwards faster than the shock, and ultimately be detected by distant observers. The same scenario holds for other types of particles, so a shock wave would release a variety of astroparticles. For a discussion of possible observational consequences in a related (but not identical) scenario, see \cite{Barrau:2014hda, Barrau:2014yka, Barrau:2015uca}.

Another possibility that we mention is that the shock wave might also generate significant gravitationally induced particle production that might be detectable.

However, the prospect for such potential phenomenological consequences could be suppressed (perhaps significantly) by the gravitational redshift of the exterior Schwarzschild geometry: if a photon escapes from the shock wave after the shock has emerged a short distance $\delta$ outside its Schwarzschild radius $R_S$, the photon's frequency would be red-shifted by a factor $\sim \delta / R_S$ (assuming $\delta \ll R_S$); this redshift could make observations challenging even if the photon is initially highly energetic.

\subsection{Information loss problem}

The prediction that the black hole lifetime is of the order $T \sim M^2 / m_{\rm Pl}$ has important implications for the black hole information loss problem.

It is well known that a black hole emits thermal radiation at the Hawking temperature $T_H = \hbar / 8 \pi G k_B M$. Assuming a quasi-static evaporation process, if black hole evaporation is the only quantum effect, its lifetime would be $\sim M^3/m_{\rm Pl}^2$ \cite{Hawking:1975vcx}. This leads to the black hole information loss problem: an initially pure state of matter on a black hole background evolves into a thermal state with non-zero entropy.

This problem with unitarity arises well before a black hole has completely evaporated. Assuming that the entropy of black hole thermodynamics $S = A/4G\hbar$ represents (the logarithm of) the number of microscopic degrees of freedom constituting a black hole, if the initial state is pure and the evolution is unitary, then any Hawking radiation (although it may appear to be thermal) must be entangled with geometric degrees of freedom in the black hole. As the black hole evaporates, the number of Hawking quanta increases while the number of black hole geometric degrees of freedom decreases. The increase in the number of Hawking quanta indicates that the required entanglement between Hawking radiation and the black hole degrees of freedom must also increase. But this is in tension with the decreasing number of black hole degrees of freedom. The Page time is when $\exp(A/4G\hbar)$, the number of degrees of freedom in the (evaporating) black hole, is no longer sufficient for the Hawking quanta to be entangled with the black hole \cite{Page:1993df, Page:2013dx}. It is at the Page time that the information loss problem truly denotes a potential loss of unitarity in the dynamics; before the Page time it is (at least in principle) possible for all Hawking radiation to be entangled with the black hole and for the total system (black hole and Hawking radiation) to remain in a pure state. The Page time is approximately half the evaporation lifetime of a black hole, and therefore is of order $\sim M^3 / m_{\rm Pl}^2$.

There are several differences between the features derived here as compared to the standard treatment. Two particularly important differences are the absence of a singularity and the absence of an eternal event horizon. In the standard semi-classical treatment of Hawking evaporation, the background space-time is assumed to be the classical Schwarzschild geometry, with a singularity and event horizon. These features make it difficult to see how information could escape a black hole: firstly, any information inside the horizon would eventually hit the singularity and be destroyed there, and (assuming energy conditions hold) information would not be able to travel in a spacelike fashion to escape the event horizon. In contrast, in the scenario we study, quantum gravity effects resolve the singularity and replace the eternal event horizon by a long-lived but temporary apparent horizon. As a result, the two obvious obstructions to information recovery are removed by quantum gravity effects, which also significantly change the causal structure of the space-time. This is exhibited in the conformal diagram in Fig.~\ref{figure:cdiag}.

This result is in agreement with earlier work in LQG, which similarly suggests that quantum gravity effects will resolve the singularity, thereby enlarging the space-time and also replacing the event horizon by an apparent horizon which eventually vanishes \cite{Ashtekar:2005qt, Perez:2014xca, Perez:2017cmj, DAmbrosio:2020mut}. However, there are some important differences in the specific way this occurs in comparison to earlier proposals, in large part due to the presence of matter. During the collapse process, an inner horizon forms inside the dust ball, and eventually exits the dust ball just before the bounce. (As a concrete example, in the Oppenheimer-Snyder collapse model the location of the inner horizon is at its `Hubble radius' $r_H = L/\dot L$, and it exits the star when $r_H = L$.) Continuity requires that this inner horizon extend into the vacuum region outside the dust ball, and this modifies in some important aspects the general scenario proposed in Refs.~\cite{Ashtekar:2005qt, Perez:2014xca, DAmbrosio:2020mut}.

Specifically, with an inner horizon, the radial coordinate $x$ (of the areal gauge) is space-like (and therefore $x=$ constant surfaces are timelike) in the region between the origin and the inner horizon, including a neighbourhood outside the dust sphere at the bounce. Due to this fact, it is impossible to glue a white hole solution to the future of a black hole solution, a scenario that requires $x=$ constant surfaces be spacelike. Hence, instead of a black hole to white hole transition, the future of the black hole in this scenario is an outgoing shock wave, as described in detail here with the conformal diagram in Fig.~\ref{figure:OS-Diag}.

In addition to the absence of a singularity and the absence of an eternal event horizon, Hawking radiation can continue only while the black hole outer horizon exists---this is the black hole lifetime $T \sim M^2 / m_{\rm Pl}$ before the shock wave reaches this horizon. (This feature is another important difference with earlier studies of black holes in LQG, which typically assume that the black hole will eventually fully evaporate after a lifetime of $T \sim M^3 / m_{\rm Pl}^2$ \cite{Ashtekar:2005qt, Perez:2014xca, DAmbrosio:2020mut}, although see \cite{Rovelli:2014cta, Haggard:2014rza} for a discussion on different possible lifetimes, including $T \sim M^2 / m_{\rm Pl}$.) Due to the shorter lifetime of the outer horizon that we find, Hawking evaporation would end well before the Page time (assuming the initial black hole mass satisfies $M \gg m_{\rm Pl}$). As a result, the amount of Hawking radiation would be relatively small (compared to what is predicted by the standard semiclassical calculations based on the assumption that the black hole completely evaporates due to Hawking radiation). Therefore, this limited amount of Hawking radiation could remain entangled with the degrees of freedom of the black hole without loss of unitarity, and information could escape with the shock wave, in gravitational and matter degrees of freedom, as the shock wave exits the horizon.

To summarize, quantum gravity effects captured by our model (i) remove the singularity, (ii) remove the event horizon, and (iii) predict the lifetime of a quantum black hole to be $\sim M^2 / m_{\rm Pl}$. The combination of these three ingredients suggests that Hawking radiation lasts for the duration of the black hole's lifetime $T \sim M^2 / m_{\rm Pl}$. As a consequence, Hawking radiation can remain entangled with black hole degrees of freedom (or perhaps with Planckian geometric or pre-geometric degrees of freedom \cite{Perez:2014xca}), and information can be recovered (at least in principle) by an outside observer once the shock wave exits the outer apparent horizon at the end of the black hole's lifetime.

We leave for future work an extension of this model to include Hawking radiation, and develop further the resolution of the information loss problem suggested here.

\newpage

\section{Summary}
\label{s.discussion}

This work provides a model for black hole formation and subsequent evolution based on a loop quantization of Lema\^itre-Tolman-Bondi space-times. Using the effective equations derived from the quantum theory, the singularity is replaced by a non-singular bounce, and a shock forms after the bounce in the gravitational field, with a discontinuity in the metric. We find weak solutions to the effective dynamics, these include analytic solutions for the Oppenheimer-Snyder and thin shell models, and numerical solutions for a variety of initial dust energy density profiles. The black hole lifetime, predicted to be $T \sim M^2 / m_{\rm Pl}$, together with the absence of a singularity or an event horizon, suggests that the information loss problem is avoided; Hawking radiation remains entangled with the black hole degrees of freedom, and ends as the shock wave exits the outer apparent horizon in a time much less than the Page time.

Our results provide a step towards providing an all-encompassing view of black hole physics---an effective quantum dynamics problem in a field theoretic setting that describes gravitational collapse to black hole formation to post-bounce dynamics. This field-theoretic description goes beyond ``quantizing the Schwarzschild metric" or the Oppenheimer-Snyder model, which are both systems with only a finite number of degrees of freedom.

Spherically symmetric systems with matter provide useful field theory models, and dust is the simplest form of matter. A next step would be to extend this model to include other types of matter that have non-vanishing pressure; for example, it would be interesting to consider a massless scalar field in spherical symmetry, a system that has been well studied classically---for some recent work in this direction (although it does not use the improved dynamics), see \cite{Benitez:2020szx}.

There remain a number of questions within the dust model. First, it would also be interesting to revisit the model without imposing any gauges before quantization. Second, the weak solutions we described are for the subset of LTB space-times with $E^b=x$; it would be interesting to see whether the main features we report here remain unaltered for the more general case without this condition. The physical argument that characteristics inevitably cross when a bounce occurs suggests that an outgoing shock wave will remain.

Another question concerns the effective equations themselves. The quantum gravity corrections they contain capture the fundamental discreteness that comes from the discrete spectrum of the area operator in loop quantum gravity, but they neglect quantum fluctuations. It would be interesting to look for solutions to the quantum dynamics that include quantum fluctuations as well, especially in the high-curvature regime.

Beyond the specific model proposed here, our results suggest a new perspective that a shock wave following a bounce may be a ubiquitous feature of singularity avoidance in black holes for any approach to quantum gravity, and raise the possibility that a lifetime of the order of $\sim M^2 / m_{\rm Pl}$ provides a means for solving the black hole information loss problem.

\newpage

\acknowledgments

\noindent
This work was supported in part by the Natural Sciences and Engineering Research Council of Canada. E.W.-E.~also acknowledges support from the UNB Fritz Grein Research Award.

\raggedright

\end{document}